\documentclass[twocolumn,tighten,preprint]{aastex62}
\usepackage{apjfonts}
\usepackage{graphicx}	
\usepackage{amssymb}
\usepackage{amsmath}
\usepackage{verbatim}
\usepackage{color}

\newcommand{\mbh}{\ensuremath{M_\mathrm{BH}}}
\newcommand{\ledd}{\ensuremath{L\mathrm{_{Edd}}}}
\newcommand{\lratio}{\ensuremath{L/\ledd}}

\newcommand{\etal}{et~al.}

\newcommand{\msun}{\ensuremath{M_{\odot}}}

\newcommand{\rs}{\ensuremath{r_{\rm \scriptscriptstyle S}}}
\newcommand{\pnull}{\ensuremath{P_{\mathrm{null}}}}

\def\logNlogS/{\ensuremath{\log N - \log S}}

\def\rev#1{{#1}}

\begin{document}

\title{A Uniformly Selected Sample of Low-mass Black Holes in Seyfert 1 Galaxies.
	III. Radio sources from the SKA pathfinders and beyond}

\author{Jin-Zhi~Wu}
\affiliation{Yunnan Observatories, Chinese Academy of Sciences, Kunming, Yunnan 650216, China}
\affiliation{School of Astronomy and Space Sciences, University of Chinese Academy of Sciences,
	19A Yuquan Road, Beijing 100049, China}

\author{Xiao-Bo~Dong}\thanks{E-mail: xbdong@ynao.ac.cn}
\affiliation{Yunnan Observatories, Chinese Academy of Sciences, Kunming, Yunnan 650216, China}

\author{Lei~Qian}
\affiliation{CAS Key Laboratory of FAST, NAOC, Chinese Academy of Sciences, Beijing 100012, China}

\author{Wen-Juan~Liu}
\affiliation{Yunnan Observatories, Chinese Academy of Sciences, Kunming, Yunnan 650216, China}

\author{Fu-Guo~Xie}
\affiliation{Key Laboratory for Research in Galaxies and Cosmology, Shanghai Astronomical Observatory,
	Chinese Academy of Sciences, 80 Nandan Road, Shanghai 200030, China}

\author{Bo~Peng}
\affiliation{School of Information Engineering, Southwest University of Science and Technology,
	Mianyang, Sichuan 621010, China}

\begin{abstract}
	Occupying the intermediate-mass regime of the accretion--jet parameter space,
	radio continuum emission from active galactic nuclei with black hole mass
	$M_{\rm BH} \lesssim 10^6 \msun$ (low-mass AGNs)
	is a valuable probe to the physics of relativistic jets.
    Yet the number of low-mass AGNs with radio detection is rather limited so far ($\approx 40$ in total).
	In this work we make two efforts to search for radio counterparts
	for the largest sample of optically selected low-mass AGNs.
	First, we collect counterparts from the recent data releases of SKA pathfinders
	such as LOFAR Two-metre Sky Survey (LoTSS). 
	Additionally, we deeply mine in
	Faint Images of the Radio Sky at Twenty-Centimeters (FIRST), 
	fitting the FIRST images of the optical AGNs with an elaborate procedure
	optimized to detect faint radio sources.
	We have obtained 151 radio sources (mainly from the SKA pathfinders),
	including 102 new reliable sources (S/N $\geqslant~5$)
	and 23 new candidates ($3.5~\leqslant$ S/N $<~5$).
    The majority of these new sources (119 of 125) have flux densities
    lower than the threshold of the official FIRST catalog. 
	The new sources have rest-frame 20 cm power ($P_{\rm 20cm}$) from $1.98\times 10^{20}$
	to $1.29\times 10^{23}$ W Hz$^{-1}$.
	For low-$z$ Seyfert galaxies
	$P_{\rm 20cm}$ correlates with $M_{\rm BH}$ intrinsically and positively,
	yet only marginally with Eddington ratio $\lratio$.
In terms of the \logNlogS/ relation for the expanding Universe, 
the limiting flux density for the completeness of our LoTSS sources
turns out to be \mbox{0.45\,mJy} at 1.4\,GHz;
i.e., complete to such a flux-density level
that is four times deeper than the official FIRST catalog.
	
\end{abstract}


\section{Introduction \label{intro}}

The physics (e.g., launching, acceleration, collimation and propagation)
of relativistic jets in active galactic nucleus (AGNs)
is an outstanding hard problem for decades,
and it also relates to various initial and boundary conditions of the supplied gas from the host galaxies
(see \citealt{Blandford2019} for a recent review).
This complicated problem is beyond analytic approach,
as can be clearly seen from the perspective of mathematics
in which it is a large set of intricately nonlinear, time-evolving partial differential equations.
It is also beyond numerical simulations and laboratory experiments in the near future,
particularly considering the two-way connections between the accretion flow and jet
(see \citealt{Davis-Tchek2020} and \citealt{Blackman-Lebedev2022} for recent reviews).
Yet progresses have been continually made over the decades;
based on the inter-fertilization of physics, numerical simulations and particularly observations,
theoretical models and new concepts have been kept improving and emerging
(see the above three reviews).


A natural way to make further progresses, from the standpoint of astronomical observations,
is to open up and explore new territory (or even new dimensions) in ``parameter space''.
As far as AGN jets are concerned,
a less explored space is radio AGNs in the low-mass end:
for the time being
the detections of radio emission
in AGNs with BH mass $\mbh \lesssim 10^6\,\msun$ (low-mass AGNs or called IMBH AGNs
\footnote{\,Following \citet{GH07sample} and \citet{Dong2012},
	hereinafter we refer to BHs with
	$\mbh < 2 \times 10^6$~\msun\ at the centers of galaxies as ``low-mass'' or
	``intermediate-mass'' BHs (\mbox{IMBHs});
	accordingly, for the ease of narration wherever it is not ambiguous,
	hereinafter we call AGNs hosting low-mass BHs
	as low-mass AGNs or IMBH AGNs.
	In the context concerning BHs yet without the term ``AGN'' occurring,
	normally we prefer ``IMBHs'' to ``low-mass BHs''
	because of the possible confusion of the latter with the stellar-mass BHs in low-mass X-ray binaries (LMXBs),
	let alone the use in the literature of ``low-mass BHs'' in the so-called ``low-mass gap'' context
	concerning the smallest stellar BHs (see, e.g., \citealt{Yang_2018_LMBH} and the mini-review therein).})  %
are rather limited.

So far, there are only 40 low-mass AGNs that have reliable radio detections in the literature.
Previously, \cite{GH07sample} listed 11 sources with FIRST detection
in their optically selected sample of broad-line low-mass AGNs.
\cite{hyLiu2018imbh} listed 26 radio sources (including 4 sources in common with \citealt{GH07sample})
by cross-matching the optically selected low-mass AGN samples of
\cite{Dong2012} and \cite{hyLiu2018imbh}
with the official FIRST catalog.
In addition,
\cite{Qian+2018} mined 35 more radio counterparts
(with $3~<$ S/N $<~5$) in FIRST images for the \cite{Dong2012} sample;
according to the strict criterion (S/N $>~5$) of the present paper for the definition of reliable radio sources,
we regard Qian \etal's radio sources as candidates only, and thus do not count them into the 40 detections.
Besides the above small radio samples (from the FIRST survey),
there are several more IMBH AGNs that are not in the FIRST catalog but have pointed observations
(e.g., \citealt{2022MNRAS.516.6123G}).
To sum up, by now the record of distinct IMBH AGNs with reliable radio detections
numbers 40 in total,
\footnote{\label{ftn:aboutA1A2}
We list in Appendix~A the pointed observations for several sources that are included in
the parent sample of the present work (namely the optical samples of \citealt{Dong2012} and \citealt{hyLiu2018imbh});
those pointed observations provided deeper and higher-resolution data than the FIRST survey at various frequencies
(see Table~\ref{tab:A1_otherData}).
We also list the observations for the 14 radio sources that are
not included in our parent sample (Table~\ref{tab:A2_otherSources}).}
which is quite small
(cf. the number of radio sources powered by supermassive BHs generally with
$\mbh \gtrsim 10^7\,\msun$;
let alone the radio galaxies and quasars
having deep and high-resolution data conducted by say VLBI, cf. \citealt{Blandford2019})
disabling almost all statistically robust conclusions regarding radio properties in the IMBH regime.

It is not easy to expand radio sample of low-mass AGNs.
Low-mass AGNs, just like the general population of low-$z$ Seyfert galaxies,
are of the so-called ``radio quiet'' (in terms of radio loudness)
in general \citep{2000ApJ...543L.111L, 2006ApJ...636...56G, GH07sample,hyLiu2018imbh}.
After all, their radio power is expected to be low
owing to their defining intermediate \mbh\ (cf. Figure~\ref{fig:power_R_on_M-lratio-diagram}).
It is thus difficult to detect their faint radio emission
by existing radio facilities within a moderate integration time.

Fortunately, thanks to the international proposal of Square Kilometre Array
(SKA; the next generation radio telescope now being constructed),
several SKA pathfinders have been built and conducted continuum surveys
(see, e.g., \citealt{2013PASA...30...20N}).
Some of the continuum surveys have released scientific data,
which have better sensitivities than
FIRST that is the deepest radio continuum survey to date
among the finished ``all-sky'' surveys.
Particularly, the pathfinder LOw Frequency ARray (LOFAR),
recently releasing its DR2 data covering 5634 deg$^2$ for the LOFAR Two-metre Sky Survey (LoTSS),
can reach very low frequencies
from several tens to about two hundreds Mega Hertz \citep{2013A&A...556A...2V},
which profits the detection of faint radio emission of synchrotron radiation nature.
Thus, it is the very time to exploit the data of these SKA pathfinders now.

In the present work, as the first step for us to harvest from the aforementioned radio data,
we collect radio counterparts as many as possible
for the optical low-mass AGNs
in two uniformly selected samples
(\citealt{Dong2012} and \citealt{hyLiu2018imbh}; totaling 513 sources).
By now the combined sample is the largest of low-mass AGNs (N.B. with \mbh\ in the IMBH regime).
Our goal is two-fold:
(1) pushing the ``parameter space'' occupied by radio AGNs
to the extremely small end along the \mbh\ dimension,
and (2) filling the IMBH regime of the radio-AGN parameter space with much more data points.

By searching in the recent data releases of the continuum surveys of SKA pathfinders,
as well as deeply mining in FIRST images (as \citealt{Qian+2018} did),
we find 102 brand-new radio counterparts (S/N $>~5$) and 23 new candidates (S/N $>~3.5$),
about 3 times the number of previously known radio sources of
low-mass AGNs ($\approx 40$).
Besides increasing the number greatly,
a second advantage is that these newly found sources are mostly below 1\,mJy
at 1.4~GHz (the flux-density limit of the official FIRST catalog).
Our radio sample are complete down to \mbox{0.45\,mJy} at 1.4~GHz,
deeper than the limit of completeness of the FIRST catalog (\cite{White97-FIRSTCatalog})
by four times.

This paper is organized as follows.
Section \ref{sect:samples}
describes the parent sample of optical-selected low-mass AGNs,
the continuum surveys of SKA pathfinders and their recent data releases,
and our deep mining of FIRST images.
Section \ref{sect:catalogs} presents
the two catalogs of the radio sources from the SKA pathfinders
and from our deep mining, respectively.
Section \ref{sect:properties} presents
preliminary studies on radio properties
and our investigation on the completeness and depth of our radio sample
in terms of the \logNlogS/ relation.
Section \ref{sect:summary} is the summary.
Throughout the paper,
we assume a cosmology with
$H_0=70\;{\rm km\;s^{-1}\;Mpc^{-1}}$,
$\Omega_m=0.3$, and $\Omega_\Lambda=0.7$.

\section{Data and sample construction \label{sect:samples}}

We search in the large-scale radio continuum surveys for radio counterparts of
the optical low-mass AGNs in the uniformly selected samples of
\cite{Dong2012} and \cite{hyLiu2018imbh} (namely Papers I and II of this series),
in order to compile a largest radio sample of low-mass AGNs.
Specifically, we conduct our searches in the publicly released data of
the continuum surveys of SKA pathfinders (\S\ref{sect:SourcesInSKAPs}),
as well as deeply mining the images of VLA FIRST (\S\ref{sect:DeepMiningFIRST}).

\subsection{Parent sample}
Our parent sample,
of which we search for radio counterparts,
is the combination of the two optically selected samples of \cite{Dong2012} and
\cite{hyLiu2018imbh}.
\cite{Dong2012} searched for type-1 AGNs having broad H$\alpha$ line
from the SDSS DR4 spectra,
and obtained 309 low-mass AGNs
with virial BH mass in the range $8\times 10^4 < M_{\rm BH} < 2 \times 10^6 M_{\sun}$.
Their virial mass estimator and threshold of low-mass AGNs ($2 \times 10^6 M_{\sun}$)
are the same as \cite{GH07sample}.
\cite{Dong2012} aimed at so-called \emph{uniformly selected} broad-line AGNs
by elaborately designing a set of automated procedure and quantitative criteria,
and attempted to probe accretion rates (namely Eddington ratio)
as low as possible (with the resulting Eddington ratios of their low-mass AGNs
ranging from $\lesssim 0.01$ to $\approx 1$).
Using this same procedure, \cite{hyLiu2018imbh}
extended the search to SDSS DR7,
and found 204 additional low-mass AGNs.
These additional sources have $1\times 10^5 < \mbh < 2 \times 10^6 M_{\sun}$,
with a similar Eddington-ratio distribution to \cite{Dong2012}.
To sum up,
our parent sample is to date the largest sample of optically selected
low-mass AGNs, totaling 513 sources (at redshifts $z < 0.35$).
They are selected from the spectroscopic data set of SDSS DR7 Legacy survey,
with a footprint area
of 8032 deg$^2$ \citep{2009ApJS..182..543A}.
The \mbh\ values range from $8\times 10^4$ to $2 \times 10^6 M_{\sun}$,
and Eddington ratios from $\lesssim 0.01$ to $\approx 1$.

\subsection{Continuum surveys of SKA pathfinders \label{sect:SourcesInSKAPs}}
So far, several radio continuum imaging sky surveys of SKA pathfinders have
publicly released their data,
such as the LOFAR Two-metre Sky Survey
(LoTSS; \citealt{2017AA....598A.104S, 2019A&A...622A...1S, 2022A&A...659A...1S})
and its deep fields version (hereafter LoTSS-deep;
\citealt{2021A&A...648A...1T, 2021A&A...648A...2S}),
Apertif imaging survey wide-shallow tier (hereafter Apertif-shallow;
Adams et al., in prep.),  and
the Karl G. Jansky Very Large Array Sky Survey (VLASS; \citealt{2020PASP..132c5001L}).

LoTSS is for the entire northern sky conducted by
LOFAR high-band antenna working in 120--168 MHz.
Two data sets with catalogs have been released,
namely the first and second data releases (called DR1 and DR2).
The DR1 focuses on a small area of 424 square degrees in HETDEX Spring Field,
with a median sensitivity of 71 $\mu {\rm Jy \; beam^{-1}}$ at 144 MHz
and an angular resolution of 6$\arcsec$ \citep{2019A&A...622A...1S}.
The DR2 covers 5634 square degrees for two regions (called RA-1 and RA-13;
RA-13 encompassing the DR1 region),
with a median sensitivity of 83 $\mu {\rm Jy \; beam^{-1}}$ at 144 MHz
and an angular resolution of 6$\arcsec$ \citep{2022A&A...659A...1S}.

LoTSS-deep has released its first data set with the catalog recently.
It covers three fields (Bo\"{o}tes, Lockman Hole and ELAIS-N1) with a total area
about 30 square degrees,
 in which the rms noise is less than $\approx$30 $\mu {\rm Jy \; beam^{-1}}$
and the resolution is 6$\arcsec$,
with respect to the working frequency of about 150 MHz \citep{2021A&A...648A...1T,2021A&A...648A...2S, 2021A&A...648A...3K}.

Apertif-shallow is conducted by the new Phased Array Feed (PAF) receiver system
Apertif installed on Westerbork Synthesis Radio Telescope (WSRT),
and plans to survey northern extragalactic sky in the frequency range 1130--1430 MHz.
Recently it has released the images of its first data
(Adams et al., in prep.)\footnote{\scriptsize http://hdl.handle.net/21.12136/B014022C-978B-40F6-96C6-1A3B1F4A3DB0\label{apertif_dr1_doc}},
which covers about 1000 square degrees (part of its planned survey area).
In this release, only data in the frequency range 1280--1430 MHz are processed,
reaching
a median rms depth of 40 $\mu {\rm Jy \; beam^{-1}}$ and a resolution of about
$15\arcsec\times15\arcsec/\sin(\delta)$.

VLASS is a survey in three epochs,
conducted by Karl G. Jansky Very Large Array
(JVLA; the upgraded version of VLA).
Every epoch will cover the same sky (the sky north of declination $-40\arcdeg$,
about 33,885 square degrees),
 with a frequency range of 2--4 GHz,
angular resolution of 2.5$\arcsec$ and
sensitivity of about 120 $\mu {\rm Jy \; beam^{-1}}$.
The sensitivity would be reduced to 70 $\mu {\rm Jy \; beam^{-1}}$ when three
epochs are combined \citep{2020PASP..132c5001L}.
Recently VLASS has released the Quick Look images of epoch 1,
with a catalog published by Canadian Initiative for Radio
Astronomy Data Analysis (CIRADA).
The Quick Look images,
not achieving the full quality of single-epoch images,
possess mean sensitivities
about 128--145 ${\rm \mu Jy \; beam^{-1}}$ \citep{2021ApJS..255...30G}.

LoTSS DR1 and DR2, LoTSS-deep DR1,
VLASS epoch 1 Quick Look (VLASS1QL) have provided their catalogs,
we thus search in them for the radio counterparts of our parent sample
by simple positional matching with a match radius of 5\arcsec.
In the case of VLASS1QL, we limit our matches to
sources with flags \textit{Duplicate\_flag} $< 2$ and \textit{Quality\_flag} $= 0$,
which indicate non-duplicates and high reliability \citep{2021ApJS..255...30G}, respectively.
It turns out that 116 distinct sources have radio counterparts,
some of which have records in more than one catalogs; see Table~\ref{tab:sample_skaPs}.

All records listed in Table~\ref{tab:sample_skaPs} have S/N $\geqslant 5$.
For the records from the catalogs of LoTSS-DR1, LoTSS-DR2 and LoTSS-deep DR1,
the peak detection threshold is set to 5
in their pipeline code of source finder \texttt{PyBDSF}
\citep{2015ascl.soft02007M};
see \citet{2019A&A...622A...1S,2022A&A...659A...1S,2021A&A...648A...1T,2021A&A...648A...2S} for the detail.
For those from VLASS1QL, setting \textit{Quality\_flag} $= 0$ means
S/N $\geqslant 5$ \citep{2021ApJS..255...30G}.

Regarding Apertif-shallow,
only the compound-beam images have been released without primary-beam correction.
We first conduct the primary beam correction,
and then fit the images corresponding to the optical AGNs in our parent sample.
The correction and the selection of best compound-beam images for every sources
follow the procedure described by Adams \etal\ (in preparation)\textsuperscript{\ref{apertif_dr1_doc}}.
We fit the Apertif-shallow images with the same code as to fit FIRST images,
which is described in detail in \S\ref{sect:fitting_procedure} below.
It turns out that
there are six radio sources satisfying S/N $\geqslant~5$,
five of which are already found in the above surveys.

\subsection{Deep mining of VLA FIRST images \label{sect:DeepMiningFIRST}}

In addition to finding radio counterparts in the surveys of SKA pathfinders,
we also mine deeply in the VLA FIRST images to extract radio sources
fainter than the flux criteria of the official FIRST catalog
(see Footnote~21 of \citealt{wjLiu+2017}, and \citealt{Qian+2018}).
The official catalog is constructed traditionally:
firstly, detect and fit possible radio sources automatedly with software;
secondly, pick up reliable sources in terms of a set of criteria.
The official criteria are (1) the best-fit peak flux density being not less than 5 times rms;
and (2) the peak flux density being not fainter than the threshold, \mbox{1\,mJy}
(i.e., the best-fit peak flux density being not fainter than \mbox{0.75\,mJy} before accounting for
the CLEAN bias; see \S4 of \citealt{White97-FIRSTCatalog}).

FIRST survey has homogeneous sensitivity of about 130 $\mu {\rm Jy \; beam^{-1}}$
and resolution of 5.4\arcsec\;at 1.4 GHz
over the vast of its coverage of 10,575 ${\rm deg}^2$;
the pixel size of the images is 1.8\arcsec\  
\citep{Becker1995_FIRST,White97-FIRSTCatalog,2015ApJ...801...26H}.
It covers nearly all the sources of our parent sample,
while LoTSS-DR2, LoTSS-deep or Apertif-shallow cover half at most.
Although VLASS (\mbox{epoch 1}) has a larger coverage area than FIRST,
the sensitivity of the current data (namely VLASS1QL)
distributes non-homogeneously \citep{2021ApJS..255...30G} and gets
worse than FIRST in some sky regions.
Thus, currently FIRST still has its advantage
in terms of both survey area and sensitivity.

We take two measures to make full use of FIRST images:
(1) an fairly elaborate procedure is applied during fitting the radio images of the optical AGNs,
with greater thoroughness and more steps than \cite{Qian+2018},
to handle correlated errors and
systematical biases or defects
(such as residual structures and even sidelobe patterns of faint sources;
\citealt{radio_book_2017,White2007_noisy});
(2) an flux criterion (S/N $\geqslant 3.5$) is carefully selected for plausible radio sources,
\footnote{In the present paper, we call the sources with $3.5 \leqslant$ S/N $< 5$ \textit{candidates}.}
after being tested and assessed quantitatively.
The two measures are described in detail below
(\S\ref{sect:fitting_procedure} and \S\ref{sect:sn-criterion}).

We would like to point out that,
to detect a possible radio source at a specific position in the sky,
a criterion of S/N less than 5 is not unreasonable
\emph{when it has been already known that there is a broad-line AGN detected in the optical}
(see also \citealt{White2007_noisy}).
We can understand this point from two perspectives.
Statistically, for a source, assuming random Gaussian noise%
\footnote{Certainly, radio astronomers concerns also the non-Gaussian errors,
which are the very subject to handle carefully in \S \ref{sect:fitting_procedure}; see the above measure (1). }
the chance probability of false detection associated with the S/N $\geqslant\;5$ criterion
is \mbox{$1-\int^{5\sigma}_{-\infty} G(x)\,\mathrm{d} x$} $=$ \mbox{3E-7},
which is of course reliable strictly.
Here $G(x)$ is the Gaussian probability density function,
zero-centered and with a standard deviation $\sigma$.
As for S/N $\geqslant\;3.5$, that probability is \mbox{2E-4},
which is actually still reliable statistically.
More important is to think in a Bayesian way:
the construction of the official FIRST catalog is essentially
a blind-source detection procedure, without any prior information (existent fact) for a specific sky position;
instead, in our case for every position in investigation
there exists an optical broad-line AGN.
It is well known that low-$z$ AGNs emit radio (at centimeter wavelengths)
more or less (cf. \S5.2 of \citealt{2008ARA&A..46..475H}).

\subsubsection{The fitting procedure \label{sect:fitting_procedure}}

We develop a fairly elaborate, interactive procedure to fit the 513 FIRST cutout images
centering at positions of the optical sources.
As described in detail by \cite{White2007_noisy},
faint sources are generally below the official $5\sigma$ threshold
(0.75 mJy, without accounting for the CLEAN bias)
and therefore susceptible to those systematical errors mentioned above,
as well as being blurred by rms noise.

First of all, we visually inspect the radio images with the corresponding optical images overlaid
to exclude those with potential contamination of background sources,
characteristic sidelobe patterns, and spot- or ridge-like structures
(see \citealt{White97-FIRSTCatalog,White2007_noisy,radio_book_2017}
for the detailed descriptions of those features).
The images of 124 sources are left, passing to the fitting in next step.

Our fitting code, based on Python
MPFIT package\footnote{https://github.com/segasai/astrolibpy/tree/master/mpfit},
is used to fit every image interactively.
First of all, we note that
during the FIRST image processing the CLEAN procedure stops
once the CLEANing limit of 0.5 mJy is reached (\citealt{Becker1995_FIRST,White2007_noisy}),
and therefore the positive and negative sidelobe patterns of faint sources remain in the images.
Thus we must include a whole region for each source,
rather than only an ``island'' of bright pixels as the customary strategy
used in pipelined softwares to fit bright radio sources (e.g., PyBDSF).
Besides, we must visually check the fitting regions to avoid serious contamination
caused by the aforementioned systematical errors and defects,
and determine a fitting region interactively on the image.
The selection procedure of the fitting region is iterated until
the histogram of flux densities of the enclosed pixels
(excluding the central pixels in the case of bright sources)
is close to a Gaussian distribution (``bell''-like shape).

We fit an elliptical Gaussian model to every fitting region.
Most of our targets have a single fitting region and thus one Gaussian component,
while some other targets have multiple fitting regions and thus multiple Gaussians.
From the best-fit Gaussian models, we derive radio parameters:
peak flux density $F_p^{\prime}$,
integrated flux density $F_i^{\prime}$,
right ascension and declination (RA and DEC; J2000),
undeconvolved position angle of major axis (measured east of north),
undeconvolved major and minor axes (FWHM).
The peak and integrated flux densities are corrected for CLEAN bias or
a similar bias
(called ``snapshot bias'' by \citealt{White2007_noisy}),
according to the empirical bias correction formula
(Equation~1 of \citealt{White2007_noisy}; also cf. \citealt{White97-FIRSTCatalog}).
We calculate the S/N of each Gaussian component to be $F_p^{\prime}/{\rm rms}$
(see the FIRST catalog description\footnote{http://sundog.stsci.edu/first/catalogs/readme.html}).
For the multi-component sources, we take the S/N of their brightest components as their S/N.

Of the above 124 sources, 86 thus fitted turn out to have S/N $\geqslant~3$
(and 33 having S/N $\geqslant~5$),
and are kept for our consideration.
Their S/N histogram is illustrated in the top panel of Figure~\ref{fig:SNR_matchratio},
where the cumulative number $N(\mathrm{S/N} > x)$ versus S/N (the $x$-axis) is plotted.
Because there are not many sources with $\mathrm{S/N} \geqslant 4$,
in the figure we use bigger incremental intervals for the range $4 \leqslant \mathrm{S/N} < 5$
than for the range $3 \leqslant \mathrm{S/N} < 4$.

\subsubsection{S/N criterion to detect deep FIRST sources \label{sect:sn-criterion}}

As stated in the beginning of \S\ref{sect:DeepMiningFIRST},
as the second measure we need to choose an appropriate S/N threshold to deeply mining
reliable sources and candidates out of the FIRST images.
The idea is this:
if we know the numbers of true and false detections versus S/N thresholds,
we can pick up an appropriate threshold value that maximizing the number of true detections
at a small cost (i.e., with a small fraction of false detections).

\begin{figure}[htb]
    \centering
    \includegraphics[width=0.4\textwidth]{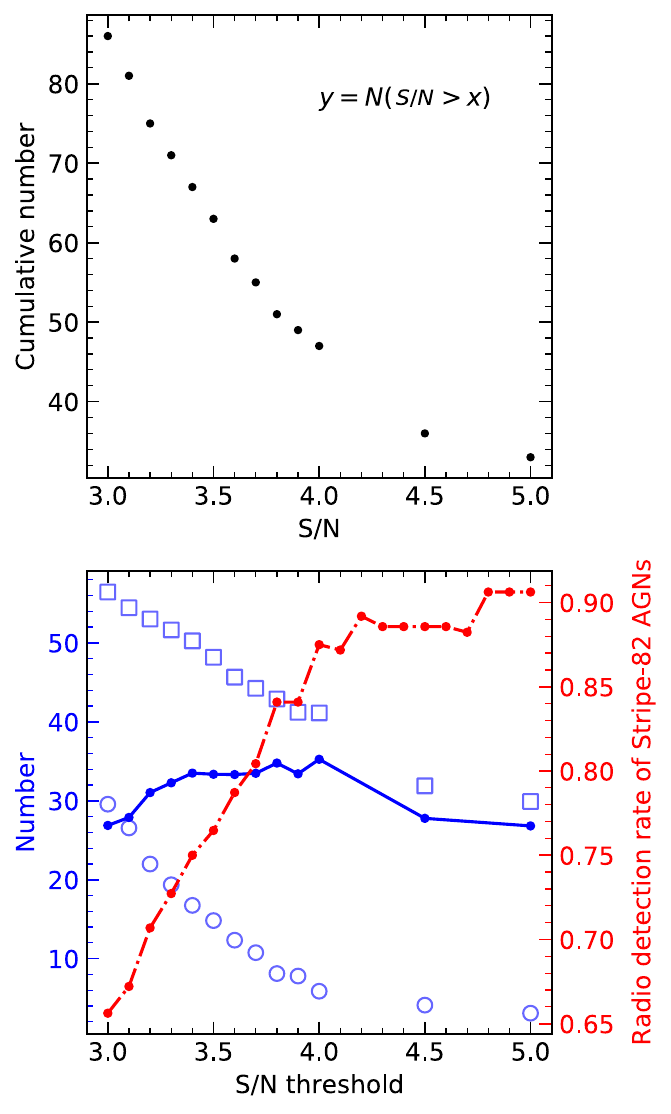}
    \caption{Determination of the S/N threshold for the radio sources deeply mined from FIRST images.
    Top panel: Cumulative histogram of the S/N values
    of the 86 FIRST counterparts for the parent sample with $\mathrm{S/N} \geqslant 3$.
    Bottom panel:
    Corresponding to the right $y$-axis (red) is
    the radio detection-rate curve of Stripe-82 AGNs (red dash-dotted line);
    the S/N values (related to $x$-axis) are based on the fitting of their FIRST images,
    and the ``true detections'' and detection rate is defined in terms of the deep VLA-S82 catalog
    (see \S\ref{sect:sn-criterion} for the detail).
    Corresponding to the left $y$-axis (blue) are
    the  estimated numbers of true (blue squares) and false detections (blue circles),
    and the number difference (the true minus false; blue solid line),
    at various S/N thresholds
    for the 86 FIRST counterparts displayed in the top panel.}
    \label{fig:SNR_matchratio}
\end{figure}

To this purpose,
we exploit the broad-line AGNs in the footprint of Deep VLA Observations of SDSS
Stripe 82 (\citealt{2011AJ....142....3H}; hereafter VLA-S82).
VLA-S82 has sensitivity and resolution three times better than FIRST.
The broad-line AGNs are from the largest broad-line AGN sample
(\citealt{hyLiu2019dr7All}) covering the Stripe 82 region.
We apply our fitting procedure to the FIRST images of these Stripe-82 AGNs.
If a thus-fitted FIRST source is recorded in the VLA-S82 catalog,
we regard it as true detection; otherwise, a false detection.
We can sense that this is a conservative definition of true detections.
The detection rate is defined as the ratio of the ``true detections'' number confirmed by VLA-S82
to the FIRST source number above a specific S/N threshold.
By setting the S/N threshold starting from $3$ progressively to $5$,
we obtain the detection-rate curve as a function of S/N threshold, 
shown in the bottom panel of Figure~\ref{fig:SNR_matchratio}
(red dash-dotted line corresponding to the right $y$-axis; all in red).

Then, we apply this detection-rate curve of Stripe-82 AGNs
to the 86 (probable) radio sources of our parent sample
(deeply mined from the FIRST images with S/N $\geqslant 3$; see \S\ref{sect:fitting_procedure}),
to estimate the numbers of true and false detections
at different S/N thresholds.
The estimated (i.e., predicted) numbers are shown
in the bottom panel of Figure~\ref{fig:SNR_matchratio},
in the same cumulative way as the top panel
(squares and circles, respectively, corresponding to the left $y$-axis; all in blue).
For the ease of devising a reasonable threshold,
we also plot the curve (blue solid line) that is the number of true detections minus that of false detections.
This blue line increases with increasing threshold (as expected),
and gets flat around S/N $=~3.5$.
Thus we choose S/N $=~3.5$ as the threshold to detect deep FIRST sources.

\begin{figure*}[htb]
  \centering
  \includegraphics[width=1\textwidth]{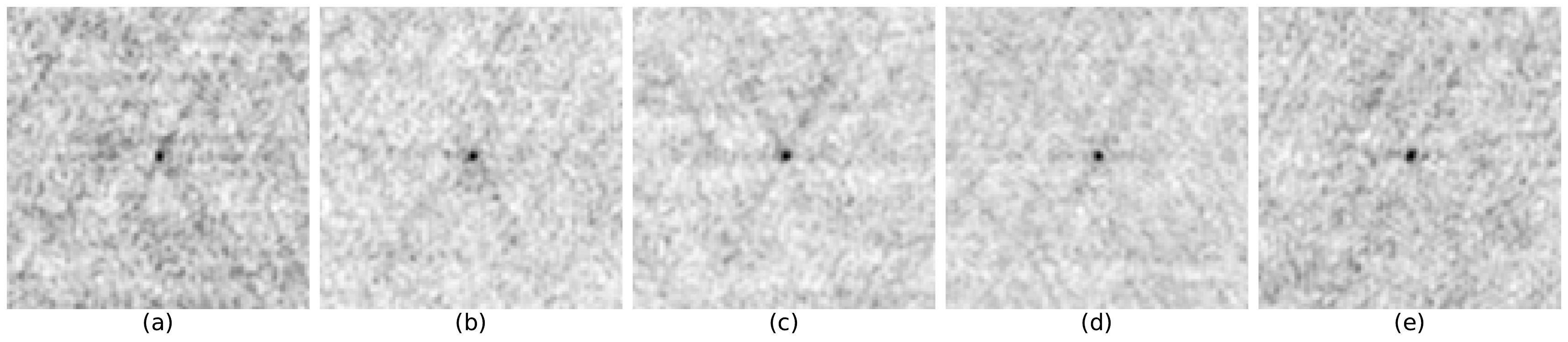}
  \caption{Stacked images for the 53 FIRST sources with $3 \leqslant \mathrm{S/N} < 5$.
  	From (a) to (e) are the five groups:
  $3 \leqslant {\rm S/N} < 3.17$ (11 sources),
  $3.17 \leqslant {\rm S/N} < 3.50$ (12),
  $3.50 \leqslant {\rm S/N} < 3.80$ (12),
  $3.80 \leqslant {\rm S/N} < 4.38$ (12), and
  $4.38 \leqslant {\rm S/N} < 5$ (6).
  The images are of size 3\arcmin$\times$3\arcmin.}
  \label{fig:stack_img}
\end{figure*}

For safety, we make a further check of the 53 FIRST counterparts with $3 \leqslant {\rm S/N} < 5$
of our parent sample,
by employing image stacking  (cf. \citealt{White2007_noisy}).
We divide those sources into five groups:
$3 \leqslant {\rm S/N} < 3.17$ (11 sources),
$3.17 \leqslant {\rm S/N} < 3.50$ (12),
$3.50 \leqslant {\rm S/N} < 3.80$ (12),
$3.80 \leqslant {\rm S/N} < 4.38$ (12), and
$4.38 \leqslant {\rm S/N} < 5$ (6).
The five stacked images based on the FIRST images
 are shown in Figure \ref{fig:stack_img}.
The S/N of their central region are calculated, which are 8.28, 9.93, 10.16,
12.21 and 10.38, respectively (from left to right).
The significant central emissions in these stacked images (even of the $3 \leqslant {\rm S/N} < 3.17$ group)
support the faint yet genuine existence of, at least a considerable fraction of, the radio sources.

\subsubsection{Accuracy of the best-fit parameters \label{sec:test_accuracy}}

We check the accuracy of our fitting
by comparing the best-fit parameters with certain references.
To quantify these comparisons, we use two indices: absolute difference $p_{\rm our} - p_{\rm ref}$,
and relative difference $(p_{\rm our} - p_{\rm ref})/p_{\rm ref}$,
where $p$ stands for any parameters such as $F_{\rm p}$, $F_{\rm i}$ and so on.

First, we conduct comparison with the official FIRST catalog.
This comparison is based on the same data; on the other hand, it is limited to bright sources
(only the 26 previously known sources, which is recorded in the official catalog, can be used).
The comparison results are presented in Table~\ref{tab:compdiff}.
The absolute differences of all parameters are small:
the mean differences are close to zero, and the standard deviations are far less
than their reference parameter values.
Relative differences are also small, with all the mean values less than 2\%
and standard deviations less than 4\%.

\begin{deluxetable*}{ccccccccccc}
	\tablewidth{0pt}
	\tabletypesize{\scriptsize}
	\tablecaption{Statistics of absolute and relative differences of radio parameters \label{tab:compdiff}}
	\centering
	\tablehead{
		\colhead{} &\colhead{}  &\multicolumn{4}{c}{Absolute differences} &\colhead{}  &\multicolumn{4}{c}{Relative differences}  \\ \cline{3-6}  \cline{8-11}
		\colhead{} &\colhead{} &\colhead{Mean} &\colhead{Std.Dev.} &\colhead{Max} &\colhead{Min} &
		            \colhead{} &\colhead{Mean} &\colhead{Std.Dev.} &\colhead{Max} &\colhead{Min}
	}
	\startdata
	R.A./arcsec & & -0.00287 & 0.05191 & 0.13489 & -0.15768 & & \nodata & \nodata & \nodata & \nodata  \\
	Dec./arcsec & & -0.00456 & 0.07093 & 0.1631 & -0.23634 & & \nodata & \nodata & \nodata & \nodata  \\
	$F{\rm_p}$/mJy/beam & & 0.01512 & 0.03474 & 0.1441 & -0.0162 & & 0.01082 & 0.02284 & 0.07624 & -0.01573  \\
	$F{\rm_i}$/mJy & & -0.03369 & 0.08301 & 0.0863 & -0.30043 & & -0.017 & 0.03781 & 0.04169 & -0.11553  \\
	Majax/arcsec & & -0.12093 & 0.29958 & 0.2983 & -1.0412 & & -0.01641 & 0.03872 & 0.03593 & -0.12918  \\
	Minax/arcsec & & -0.07475 & 0.16834 & 0.1532 & -0.6344 & & -0.01266 & 0.02639 & 0.02170 & -0.08848  \\
	P.A./degree & & -1.85025 & 8.87480 & 8.2801 & -35.2566 & & \nodata & \nodata & \nodata & \nodata  \\
	\enddata
	\tablecomments{This table lists the mean, standard deviation, maximum and minimum of absolute and relative differences of seven radio parameters, between our best-fit parameter values and those in the official FIRST catalog of the 26 known sources.
	}
\end{deluxetable*}

Second, we conduct comparison by exploiting the small survey VLA-S82 (see \S\ref{sect:sn-criterion}),
to test the accuracy of our fitting to the FIRST images of faint sources ($3 \leqslant \mathrm{S/N} < 5$).
For this purpose, the 10 broad-line AGNs in \S\ref{sect:sn-criterion},
which have $3 \leqslant \mathrm{S/N} < 5$ fits based on the FIRST images
and are recorded in the VLA-S82 catalog (noting than VLA-S82 is deeper than FIRST),
are also used here.
Because the spatial resolution is different between VLA-S82 and FIRST and the sample size is small,
we only consider $F{\rm_p}$ (the directly fitted and important quantity).
The mean of the $F{\rm_p}$ difference is $-0.041$ mJy/beam;
the standard deviation is 0.104 mJy/beam.
Both are smaller than the rms noise of FIRST images (0.14 mJy/beam median).


\section{New catalogs \label{sect:catalogs}}

As described in the preceding section,
we obtain two radio subsamples of low-mass AGNs.
In the subsample from the surveys of SKA pathfinders (SKAPs subsample in short),
there are 117 \emph{reliable sources} (S/N $\geqslant~5$), among which 100 are brand-new.
In the subsample deeply mined from the FIRST images (deep FIRST subsample in short),
there are 63 sources (S/N $\geqslant~3.5$),
among which 33 are deemed as reliable sources with S/N $\geqslant~5$
(including the 26 previously known sources recorded in the official FIRST catalog),
and the remaining 30 are deemed as \emph{candidate sources} with $3.5 \leqslant \mathrm{S/N} < 5$.
We compile two respective catalogs,
as presented in Table~\ref{tab:sample_skaPs} and Table~\ref{tab:sample_deepFIRST}.
There are 29 sources in common between the two subsamples.
Combining the two subsamples, there are 151 distinct sources in total,
including the 26 known sources, 102 reliable sources, and 23 candidate sources.

In addition, considering that LoTSS-DR2 is both deep and wide
(now covering about half of the sky area of our parent sample),
we present upper limits for the 104 low-mass AGNs that are covered by LoTSS-DR2 yet under detection.
The upper-limit flux density is set to be 5 times the local rms of each source,
where rms is from the official rms map of LoTSS-DR2.
The data of these upper-limit sources are placed in Appendix~\ref{appendix:LoTSS-upperlimits}
(see Table~\ref{tab:upperlimit_sources}).
Applications of these sources are deferred to future work;
in this study, we only display the upper-limit distributions of their radio power and loudness
(in Figures \ref{fig:Lhist} and \ref{fig:Rhist}) as reference.
We do not consider upper limits based on the FIRST data
since it is relatively shallow now in this SKA-pathfinder era.

\begin{longrotatetable}
  \begin{deluxetable*}{cccccccccccccccccccc}
    \tablewidth{0pt}
    \tabletypesize{\tiny}
    \tablecaption{Radio low-mass AGNs from the surveys of SKA pathfinders \label{tab:sample_skaPs}}
    \tablehead{
      \colhead{ID} &\colhead{SDSS Name} &\colhead{DR} &\colhead{Freq} &\colhead{P} &\colhead{RA} &\colhead{RA\_e} &\colhead{DEC} &\colhead{DEC\_e} &\colhead{$F_{\rm p}$} &\colhead{$F_{\rm p\_e}$} &\colhead{$F_{\rm i}$} &\colhead{$F_{\rm i\_e}$} &\colhead{PA} &\colhead{PA\_e} &\colhead{MajAx} &\colhead{MajAx\_e} &\colhead{MinAx} &\colhead{MinAx\_e} &\colhead{Note} \\
      \colhead{} &\colhead{} &\colhead{} &\colhead{MHz} &\colhead{} &\colhead{deg} &\colhead{arcsec} &\colhead{deg} &\colhead{arcsec} &\colhead{mJy/} &\colhead{mJy/} &\colhead{mJy} &\colhead{mJy} &\colhead{deg} &\colhead{deg} &\colhead{arcsec} &\colhead{arcsec} &\colhead{arcsec} &\colhead{arcsec} &\colhead{} \\
      \colhead{} &\colhead{} &\colhead{} &\colhead{} &\colhead{} &\colhead{} &\colhead{} &\colhead{} &\colhead{} &\colhead{beam} &\colhead{beam} &\colhead{} &\colhead{} &\colhead{} &\colhead{} &\colhead{} &\colhead{} &\colhead{} &\colhead{} &\colhead{} \\
      \colhead{(1)} &\colhead{(2)} &\colhead{(3)} &\colhead{(4)} &\colhead{(5)} &\colhead{(6)} &\colhead{(7)} &\colhead{(8)} &\colhead{(9)} &\colhead{(10)} &\colhead{(11)} &\colhead{(12)} &\colhead{(13)} &\colhead{(14)} &\colhead{(15)} &\colhead{(16)} &\colhead{(17)} &\colhead{(18)} &\colhead{(19)} &\colhead{(20)}
          }
    \startdata
    S1 & J073106.87+392644.7 & LoTSS-DR2 & 120-168 & 1 & 112.77844 & 0.20 & 39.44568 & 0.23 & 1.262 & 0.083 & 2.067 & 0.203 & 160.15 & 23.17 & 8.15 & 0.56 & 7.24 & 0.45 & F2\\
    S2 & J073631.83+383058.4 & LoTSS-DR2 & 120-168 & 1 & 114.13288 & 0.52 & 38.51614 & 0.60 & 0.534 & 0.075 & 1.299 & 0.247 & 176.97 & 63.47 & 9.82 & 1.42 & 8.92 & 1.22\\
    S3 & J073956.02+402816.3 & LoTSS-DR2 & 120-168 & 1 & 114.98441 & 0.12 & 40.47119 & 0.13 & 2.451 & 0.103 & 3.446 & 0.225 & 165.31 & 26.29 & 7.36 & 0.32 & 6.87 & 0.28\\
    S4 & J074251.09+333403.8 & LoTSS-DR2 & 120-168 & 1 & 115.71303 & 0.08 & 33.56774 & 0.09 & 3.797 & 0.115 & 4.530 & 0.225 & 142.56 & 28.88 & 6.70 & 0.21 & 6.41 & 0.19 & F3;FCat\\
    S5 & J074948.33+264734.2 & LoTSS-DR2 & 120-168 & 1 & 117.45137 & 0.03 & 26.79282 & 0.02 & 11.205 & 0.105 & 12.719 & 0.198 & 89.09 & 6.60 & 6.59 & 0.06 & 6.21 & 0.06 & F5;FCat\\
    S5 & J074948.33+264734.2 & VLASS1QL & 2000-4000 & 1 & 117.45149 & 0.11 & 26.79288 & 0.04 & 2.269 & 0.126 & 2.374 & 0.233 & 83.09 & 5.75 & 3.69 & 0.26 & 2.28 & 0.10 & F5;FCat\\
	  \enddata
    \tabletypesize{\scriptsize}
    \tablecomments{Column (1), identification number assigned in this paper for the SKA-pathfinders sources.
    	Column (2), SDSS name.
    	Column (3), data release of the surveys of SKA pathfinders.
    	Column (4), frequency range of the surveys.
    	Column (5), the parent catalog; `1' standing for \cite{Dong2012} and `2' for \cite{hyLiu2018imbh}.
    	Columns (6) to (9), right ascension and declination (J2000) measured for the radio sources, and the positional errors.
    	Columns (10) to (13), peak and integrated flux densities, and their errors.
    	Columns (14) to (19), undeconvolved position angle (east of north), major and minor axes (FWHM), and their errors.
    	Column (20) is the additional note:
    	the sources having detection in the deep FIRST subsample are marked with their identification numbers in that subsample;
    the sources listed in the official FIRST catalog, marked with `FCat';
    the source having multiple Gaussian components, marked with `M'.\\
    (This table is available in its entirety in a machine-readable form in the online journal. A portion is shown here for guidance regarding its form and content.)}
  \end{deluxetable*}
\end{longrotatetable}


\begin{longrotatetable}
  \begin{deluxetable*}{ccccccccccccccccccc}
    \tablewidth{0pt}
    \tabletypesize{\tiny}
    \tablecaption{Expanded radio low-mass AGNs deeply mined from FIRST images \label{tab:sample_deepFIRST}}
    \tablehead{
      \colhead{ID} &\colhead{SDSS Name} &\colhead{P} &\colhead{RA} &\colhead{RA\_e} &\colhead{DEC} &\colhead{DEC\_e} &\colhead{$F_{\rm p}$} &\colhead{$F_{\rm p\_e}$} &\colhead{$F_{\rm i}$} &\colhead{$F_{\rm i\_e}$} &\colhead{$F_{\rm i\_tot}$} &\colhead{PA} &\colhead{PA\_e} &\colhead{MajAx} &\colhead{MajAx\_e} &\colhead{MinAx} &\colhead{MinAx\_e} &\colhead{Note} \\
      \colhead{} &\colhead{} &\colhead{} &\colhead{deg} &\colhead{arcsec} &\colhead{deg} &\colhead{arcsec} &\colhead{mJy/beam} &\colhead{mJy/beam} &\colhead{mJy} &\colhead{mJy} &\colhead{mJy} &\colhead{deg} &\colhead{deg} &\colhead{arcsec} &\colhead{arcsec} &\colhead{arcsec} &\colhead{arcsec} &\colhead{} \\
      \colhead{(1)} &\colhead{(2)} &\colhead{(3)} &\colhead{(4)} &\colhead{(5)} &\colhead{(6)} &\colhead{(7)} &\colhead{(8)} &\colhead{(9)} &\colhead{(10)} &\colhead{(11)} &\colhead{(12)} &\colhead{(13)} &\colhead{(14)} &\colhead{(15)} &\colhead{(16)} &\colhead{(17)} &\colhead{(18)} &\colhead{(19)}
          }
    \startdata
    F1 & J031743.12+001936.8 & 1 & 49.42972 & 0.15 & 0.32687 & 0.17 & 1.626 & 0.101 & 1.833 & 0.227 & 1.833 & 172.47 & 18.78 & 6.65 & 0.40 & 5.86 & 0.35 & FCat\\
    F2 & J073106.87+392644.7 & 1 & 112.77892 & 0.58 & 39.44571 & 0.39 & 0.709 & 0.129 & 0.770 & 0.264 & 0.770 & 78.85 & 18.96 & 6.93 & 1.44 & 4.57 & 0.92 & S1\\
    F3 & J074251.09+333403.8 & 1 & 115.71285 & 0.05 & 33.56774 & 0.05 & 4.220 & 0.213 & 3.174 & 0.202 & 3.174 & 0.00 & 0.00 & 4.70 & 0.13 & 4.67 & 0.13 & S4;FCat\\
    F21.1 & J103618.78+051958.2 & 1 & 159.07621 & 0.69 & 5.33274 & 0.49 & 0.675 & 0.145 & 0.288 & 0.184 & 1.144 & 65.00 & 41.07 & 4.27 & 2.09 & 2.91 & 1.01\\
    F21.2 & J103618.78+051958.2 & 1 & 159.07754 & 0.66 & 5.33274 & 0.67 & 0.771 & 0.145 & 0.857 & 0.560 & 1.144 & 133.05 & 59.75 & 6.71 & 3.51 & 4.83 & 1.66\\
    \enddata
    \tabletypesize{\scriptsize}
    \tablecomments{Column (1), identification number assigned in this paper for the deep FIRST sources.
    Columns (2) to (7) and (13) to (18), the same definitions as in the above Table~\ref{tab:sample_skaPs};
    see the text for the methods of deriving the parameter values.
    Column (12), the total integrated flux density,
    which is the sum of the integrated flux densities of the individual Gaussian components for multi-component sources.
    Column (19), the additional note:
    the sources having detection in the SKAPs subsample are marked with its identification number in that subsample;
    the sources in the official FIRST catalog, marked with `FCat'. \\
    (This table is available in its entirety in a machine-readable form in the online journal. A portion is shown here for guidance regarding its form and content.)}
  \end{deluxetable*}
\end{longrotatetable}

\clearpage

\subsection{Radio subsample from SKA pathfinders \label{sect:Cat_SKAPs}}
The data of the 117 radio sources in the SKAPs subsample are listed
in Table~\ref{tab:sample_skaPs}. The columns are as follows:

Column (1), identification number (ID) assigned for the radio sources in this paper.
Different detections of a source have the same ID.
Prefix `S' represents the surveys of SKA pathfinders.

Column (2), the SDSS name (JHHMMSS.ss $+$DDMMSS.s).

Column (3), data releases of the surveys of SKA pathfinders.

Column (4), frequency range of the surveys (in units of MHz).

Column (5), the parent catalog with `1' standing for \cite{Dong2012} and `2' for \cite{hyLiu2018imbh}.

Columns (6) to (9), right ascension (RA) and declination (DEC) in decimal degrees (J2000.0)
measured for the radio sources,
and their measurement errors in arcseconds;
the values are from the SKA pathfinders' catalogs
(except the Apertif sources; see below).

Columns (10) to (13), peak flux density $F_{\rm p}$ and its error in mJy beam$^{-1}$; integrated flux density $F_{\rm i}$ and its error in mJy.

Columns (14) and (15), undeconvolved position angle (PA, measured east of north) and its error, in degrees.

Columns (16) to (19), undeconvolved major and minor axes (FWHM) and their errors, in arcseconds.

Column (20) is the additional note. If a source also having a detection in the deep FIRST subsample,
we mark its identification number in that subsample;
if a source is already listed in the official FIRST catalog, marked with `FCat';
if a source has multiple Gaussian components, marked with `M'.

Regarding the parameters in Columns (6) to (19):
for the sources selected from surveys of LoTSS, LoTSS-deep and VLASS1QL,
the parameter values come from their official radio catalogs;
for the sources selected from Apertif-shallow continuum images,
the values are given by MPFIT in the 2D Gaussian fitting
(except that the error of peak flux density is the local rms noise of the images,
and the error of integrated flux density is calculated through error propagation).

\subsection{Radio subsample deeply mined from FIRST \label{sect:Cat_deep_FIRST}}

The data of the 63 radio sources in the deep FIRST subsample are listed
in Table~\ref{tab:sample_deepFIRST}.
For the sources fitted with multiple Gaussian components,
following \cite{White97-FIRSTCatalog} we list every individual components in the catalog.
The columns are as follows:

Column (1), identification number assigned in this paper,
with decimals in digit representing the individual Gaussian components of those multi-component sources.
Prefix `F' represents FIRST survey.

Columns (2) to (11) and (13) to (18), the same definitions as in \S\ref{sect:Cat_SKAPs}.
Note that the peak and integrated flux densities of the deep FIRST sources
have been corrected for CLEAN bias or ``snapshot bias'' (cf. \S\ref{sect:fitting_procedure}).

Column (12), the total integrated flux density $F_{\rm i\_tot}$ in mJy,
which is the sum of the integrated flux densities of
the individual Gaussian components for multi-component sources, following \cite{2005MNRAS.362....9B}.

Column (19), the note.
The sources also having detection in the SKAPs subsample are marked with their identification numbers in that subsample;
the sources in the official FIRST catalog, marked with `FCat'.

The parameter values in columns (4) to (18) are given by MPFIT in the 2D Gaussian fitting
(except the errors to the peak and integrated flux densities; see \S\ref{sect:Cat_SKAPs}).

In the deep FIRST subsample, the 29 sources in common with the SKAPs subsample are as follows:
17 previously known sources (in the official FIRST catalog),
5 newly found reliable sources (S/N $\geqslant~5$),
and 7 candidate sources ($3.5 \leqslant \mathrm{S/N} < 5$).
Four of the 12 new sources (5 $+$ 7) have data of pointed observations in the literature,
which are marked in  Table~\ref{tab:A1_otherData}.
We noticed that
among the 37 new FIRST sources (i.e., excluding the 26 ones recorded in the official FIRST catalog),
16 are covered by the footprints of LoTSS-DR2,
giving a confirmation fraction by LOFAR of 75\% (namely 12/16).

Except the above 63 radio candidates
out of the 86 (probable) radio sources
deeply mined from the FIRST images for our parent low-mass AGN sample
(\S\ref{sect:fitting_procedure}),
there are 23 remaining sources with $3 \leqslant \mathrm{S/N} < 3.5$.
They are optimal targets of radio low-mass AGNs for
deep and high-resolution observations in the future,
and thus we list their information (including the best-fit parameters based on their FIRST images)
in Appendix~\ref{appendix:first_3_3.5}.
\newline

\section{Sample properties \label{sect:properties}}
As stated above,
this work has obtained 151 radio low-mass AGNs,
of which 128 are reliable detections (with S/N $\geqslant 5$).
Either of the two numbers is more than 3 times
the number of known radio low-mass AGNs (totaling 40).
Of the 151 sources, 125 are newly found
(from the surveys of SKA pathfinders and our deep mining of FIRST images).
Among our newly found sources,
119 sources are fainter than the flux threshold of the official FIRST catalog
(1\,mJy at 1.4~GHz;
regarding sources from SKA pathfinders, for comparison their flux densities are converted to
those at 1.4~GHz assuming a radio spectral index of 0.46).
Thus our radio sample extends deeper into the faint-flux regime
than those previous small radio samples of low-mass AGNs (see \S\ref{intro}).
This encourages us to
investigate the statistical properties of the radio sources,
and assess the completeness and depth of the present radio sample.

We note that,
due to the insufficient spatial resolution of the radio data to identify radio emissions
from the Seyfert nuclei,
strictly speaking, these radio sources are of plausible AGN origin;
thus the results of our investigations on jet emission
are instructive but not conclusive.

\subsection{Morphology, spectral index, power and radio loudness\label{sect:mpr}}

\subsubsection{Morphology\label{sect:morphology}}

Considering the hybrid parentage and not decent spatial resolutions of these radio sources,
here we take a conservative way to identify ``extended'' radio sources:
we only classify these sources to be either single-component or multi-component,
in terms of the number of best-fit Gaussian components.
It turns out that
132 of the 151 sources have only one Gaussian component,
and the rest 19 have two or more Gaussian components.
Ten of the multi-component sources are in the SKAPs subsample
(we mark them with `M' in Table~\ref{tab:sample_skaPs}),
and nine in the deep FIRST subsample
(denoted by decimals in their identification numbers in Table~\ref{tab:sample_deepFIRST}).
The (probably) most extended sources, in terms of
the maximum separation among the Gaussian components,
have 10.9 kpc in the SKAPs subsample,
and 20.9 kpc in the deep FIRST subsample.

\begin{figure*}[htb]
  \centering
  \includegraphics[width=1\textwidth]{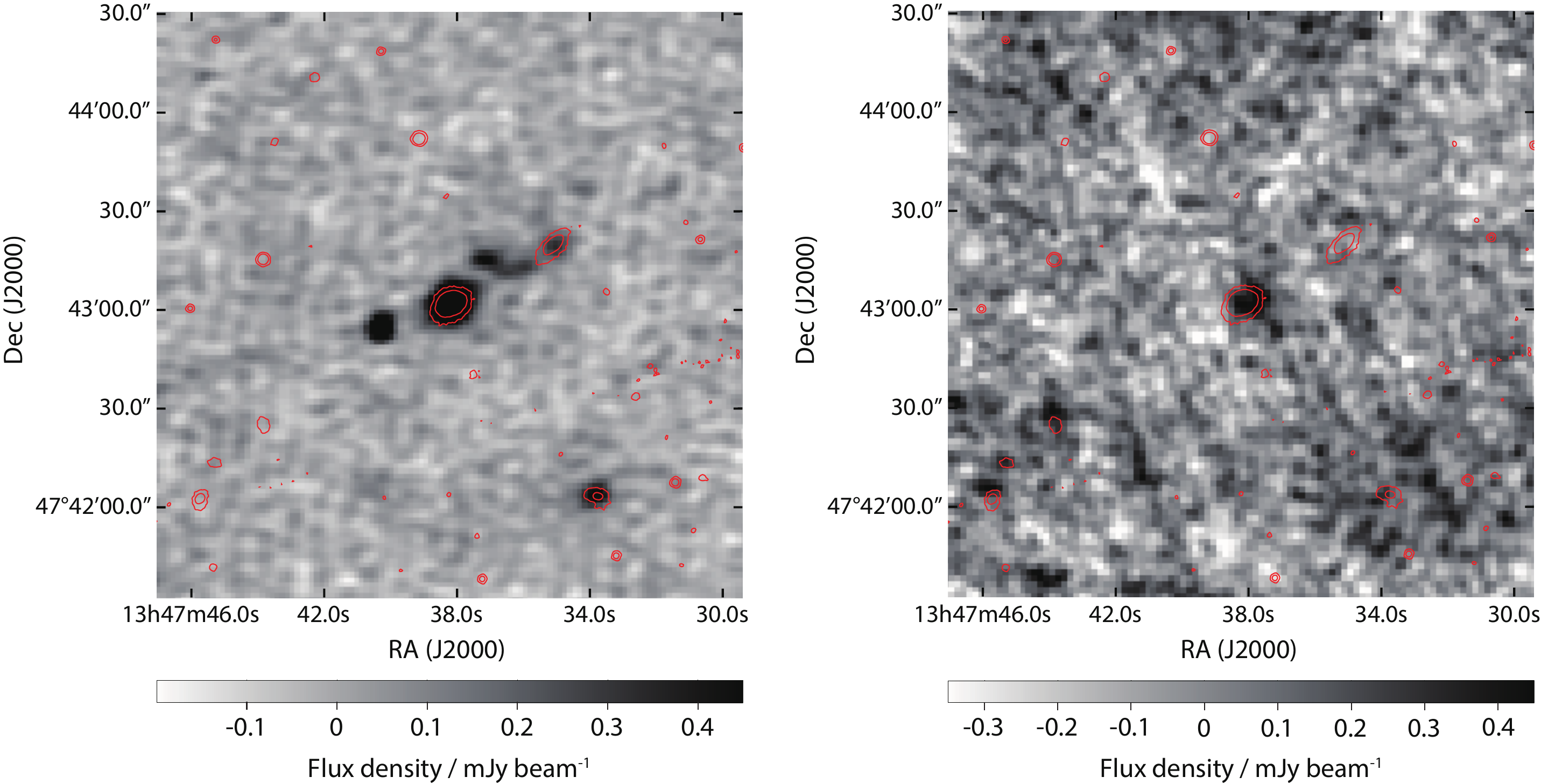}
  \caption{\rev{The LoTSS-DR2 (left) and FIRST (right) images of J134738.24$+$474301.9, 
  		with the contours of its optical image overlaid. 
  The flux density in the images are color-coded according to the respective colorbars.}
    }
  \label{fig:special_sources}
\end{figure*}

Brief notes are in order for two special multi-component sources.
J122342.81$+$581446.1 in the SKAPs subsample
has two Gaussian components,
but the optical image indicates one of the two radio components sits on a tiny background galaxy 
\rev{(J122343.72$+$581454.3 at $z = 0.588$ as identified from its SDSS image and spectrum)%
\footnote{https://skyserver.sdss.org/dr18/VisualTools/explore/summary?objId=1237657612878217246}%
}; 
it is not clear at this point if there is any connection between that radio component and the background galaxy.
Thus we mark this source with \mbox{`M?'} in Table~\ref{tab:sample_skaPs}.
The second is J134738.24$+$474301.9 in the deep FIRST subsample,
for which we need two Gaussian components to fit.
Interestingly, it has been actually recorded also in the LoTSS-DR2 catalog
yet without a multi-component tag.
We inspect its LoTSS image,
and see multiple radio sources
within a radius 28\arcsec\
(namely, a projected physical radius of 33.6 kpc)
without any other optical counterparts.
Thus the spatially distributed radio emission
is of the same source plausibly,
with the brightest component J134738.24$+$474301.9 located in the center;
all these components appear to constitute a jet, roughly in a linear configuration.
\rev{The LoTSS-DR2 and FIRST images, with its optical image from the SDSS (shown as contours) overlaid,
are shown in Figure \ref{fig:special_sources}. 
The low-mass AGNs with extended jets are very few so far, and valuable for jet physics 
(see, e.g., SDSS J095418.15+471725.1 by \citealt{2022MNRAS.516.6123G}).
Thus J134738.24$+$474301.9 deserves deep and high-resolution pointed radio observations in the future.  
}

\subsubsection{Spectral index\label{sect:spectral-index}}

There are 25 sources in our sample that have detections in both LoTSS-DR2 and deep FIRST subsamples.
\rev{There is a large frequency difference between LoTSS and FIRST, centering at 144 MHz and 1.4 GHz, respectively,	
which profits the investigation of radio spectral slope.
Moreover, the angular resolutions of the two surveys are similar (6\arcsec\ and 5.4\arcsec, respectively),
minimizing the potential bias in flux due to resolution mismatch.}
Assuming there is no considerable variability between different epochs (at least statistically),
we investigate  
\rev{the two-point radio spectral indices of the 25 sources.}
The indices are calculated by 
using their integrated flux densities
and adopting the spectral form $f_\nu \propto \nu^{-\alpha_r}$, where $\alpha_r$ is the radio spectral index.

\begin{figure}[htb]
  \centering
  \includegraphics[width=0.47\textwidth]{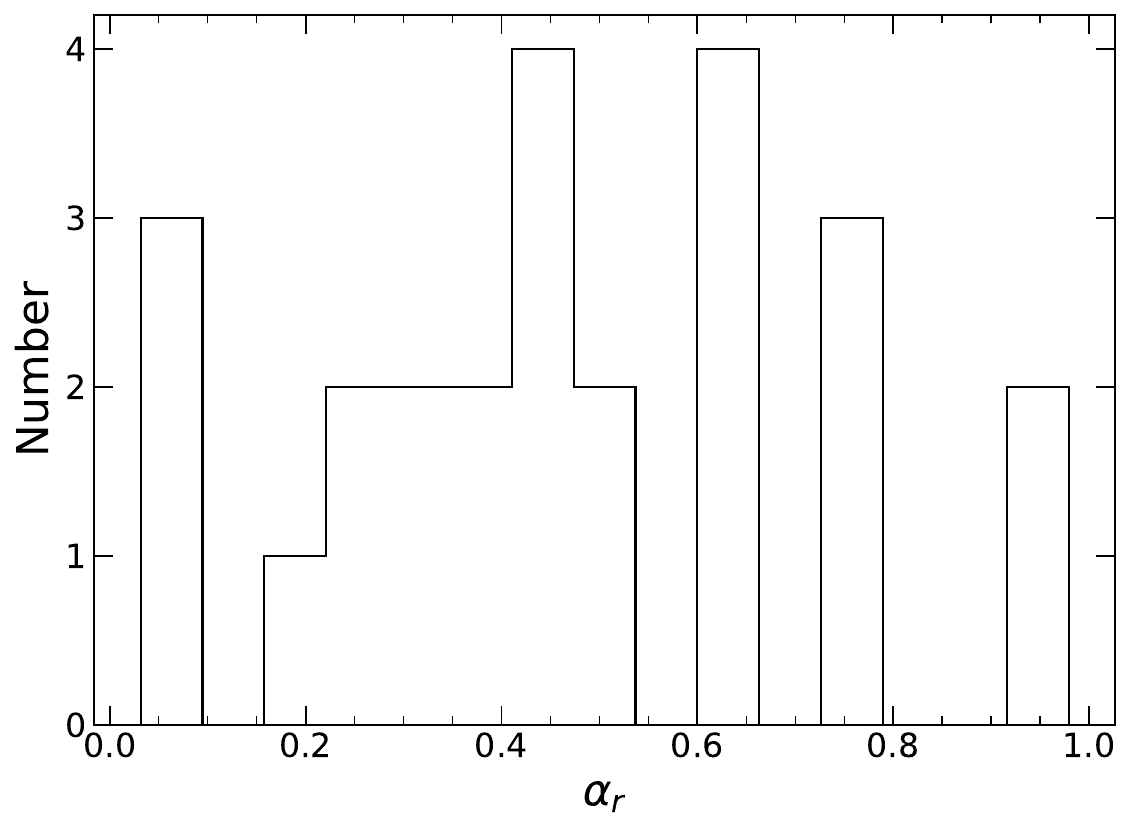}
  \caption{Histogram of the radio spectral indices of the 25 sources in both LoTSS-DR2 and deep FIRST subsamples.}
  \label{fig:specidx}
\end{figure}

As shown in Figure~\ref{fig:specidx}, the derived $\alpha _r$ values span from 0.03 to 0.98,
with a standard deviation of 0.26, a mean of 0.47 and a median of 0.45,
consistent with \cite{2001ApJS..133...77H}.
\rev{Adopting the generally used division $\alpha_r = 0.5$ 
	between flat and steep radio spectra,
14 of the 25 sources are flat ($\alpha_r < 0.5$).
In the context of AGNs, flat radio spectra are from  
compact radio cores, and steep spectra from 
extended jets or lobes 
(see \S1.3.1 of \citealt{1997iagn.book.....P}).
More importantly, while it is also possible for steep radio spectra to originate from star formation activity,
it is generally believed that flat-spectrum radio sources
cannot be associated with star formation. 
Interestingly, 5 of the 11 steep-spectrum sources are 
multi-component in morphology as described in \S\ref{sect:morphology}, 
while only 3 of the 14 flat-spectrum sources are fitted with multiple Gaussians.
With $\alpha _r$ being 0.45 and 0.37 two of them are near the boundary to be steep-spectrum,
and the third one is J134738.24$+$474301.9 with an extended jet as shown in Figure~\ref{fig:special_sources},
of which the low-frequency flux (LoTSS)  might be underestimated owing to the extended structure (see \S\ref{sect:morphology})
and thus its true spectral index could be steeper.}

\subsubsection{Radio power\label{sect:Power}}

We make quick statistics to the powers at rest-frame 20~cm  $P_{\rm 20cm}$ of the radio sources.
To calculate the powers, we use the integrated flux densities specified as follows.
For the sources having multiple detections in the SKAPs subsample,
their flux densities from LoTSS-DR2 are used;
for the 29 sources having detections in the both subsamples,
the flux densities from the SKAPs subsample are used for the 12 newly-found ones
(because their SKAPs fluxes are above 5$\sigma$),
and their FIRST flux densities are used for the 17 previously known
(measured at the rest-frame 20 cm wavelength and with S/N $\geqslant~5$);
the total integrated flux densities are used for the multi-component sources.
For flux densities at frequencies different from 1.4 GHz,
we convert them to 1.4 GHz (namely at the 20~cm wavelength) by assuming
a radio spectral index $\alpha _r=0.46$
(\citealt{2001ApJS..133...77H, 2001ApJ...558..561U}).
The $\log P_{\rm 20cm}$ values of our radio sources are recorded in Table~\ref{tab:R}.
The $P_{\rm 20cm}$ values of the 125 newly found sources range from
$1.98 \times 10^{20}$ to $1.29 \times 10^{23}\;{\rm W\; Hz^{-1}}$
with a median of $1.12 \times 10^{22}\;{\rm W\; Hz^{-1}}$,
while the $P_{\rm 20cm}$ of the 26 known sources  (i.e., listed in the official FIRST catalog)
from $2.40 \times 10^{21}$ to $3.72 \times 10^{23}\;{\rm W\; Hz^{-1}}$
with a median of $3.34 \times 10^{22}\;{\rm W\; Hz^{-1}}$.

For comparison,
we also calculate the rest-frame $P_{\rm 20cm}$ of the sample of
broad-line AGNs in SDSS DR7 (\citealt{hyLiu2019dr7All};
dubbed BLAGNs hereinafter, excluding the 26 known radio low-mass AGNs of \citealt{hyLiu2018imbh}
since they have been included in our sample).
For those BLAGNs we use their integrated flux densities from the FIRST catalog.

\begin{figure}[htb]
\centering
\includegraphics[width=0.48\textwidth]{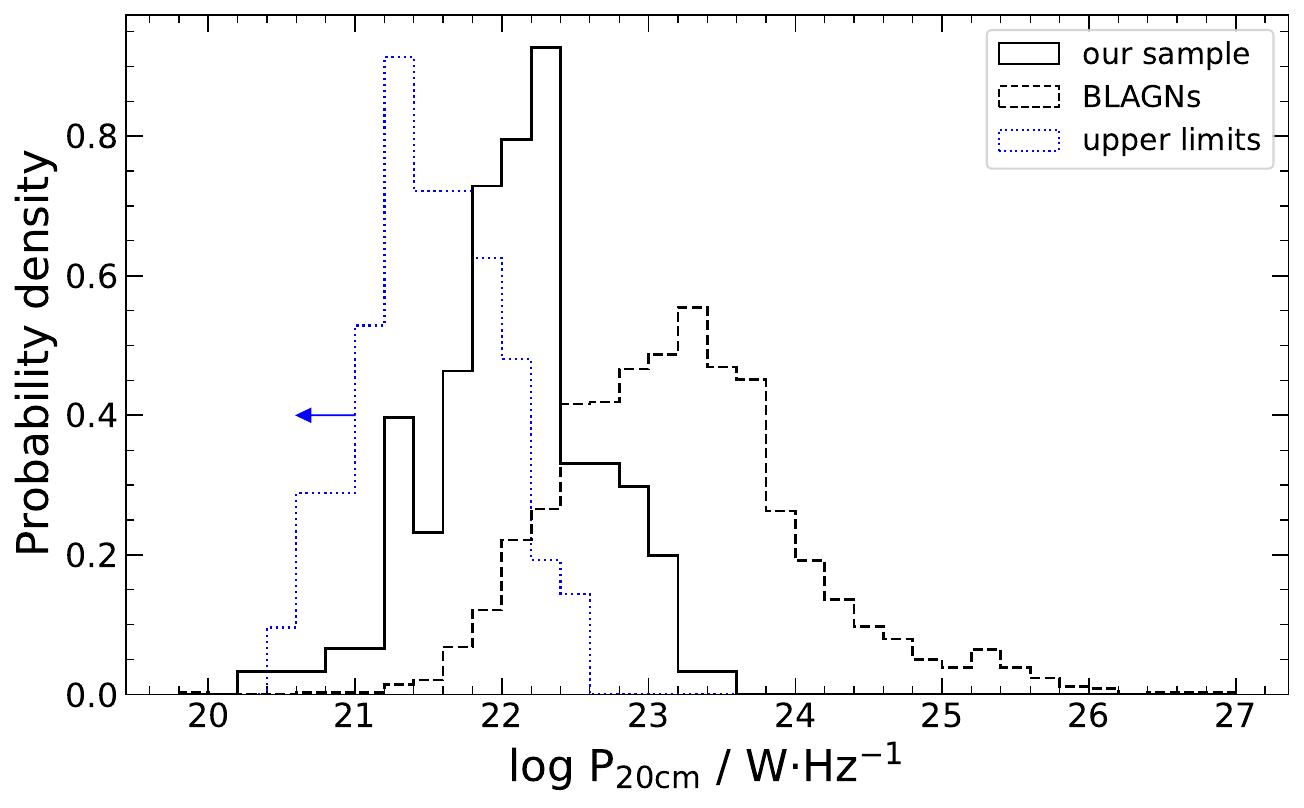}
\caption{Normalized histograms of the rest-frame power at 20 cm for our radio sample (151 sources; black solid line)
	and all the other radio sources in the SDSS-DR7 broad-line AGN sample (dubbed BLAGNs in the text; black dashed line).
	Also plotted is the distribution of the upper-limit radio sources for our parent sample of optical low-mass AGNs
	(104 sources; blue dotted line).
  The histograms are normalized to make the probability (namely the area under every histogram) being unity.}
\label{fig:Lhist}
\end{figure}

The $\log P_{\rm 20cm}$ histograms of our radio sample and the so-called BLAGNs
are plotted in Figure~\ref{fig:Lhist}.
The histograms are normalized to have unity probability (namely the area under every histogram).
The distribution of our 151 low-mass AGNs is in the lower end of the BLAGN distribution,
which has a larger median of $1.58\times 10^{23}\;{\rm W\; Hz^{-1}}$.

\subsubsection{Radio loudness\label{sect:Loudness}}

Radio loudness $R$,
an indicator of the relative intensity of jet to accretion-disk emission,
is defined as the ratio in the rest frame, $R\equiv f_\nu (6{\rm\; cm})/f_\nu (4400${\AA}).
We calculate the rest-frame $f_\nu (6{\rm\; cm})$,
by using the same integrated flux densities, radio spectral index $\alpha _r$ and
rest-frame correction as used above to calculate the rest-frame $P_{\rm 20cm}$.
The rest-frame $f_\nu (4400$\AA) is converted from
monochromatic luminosity $L_\lambda $(5100\AA),
 by assuming an optical spectral index $\alpha _o=0.44$\;($f_\nu=\nu^{-\alpha _o }$; \citealt{2001AJ....122..549V}).
The $L_\lambda $(5100\AA) data are
derived from the luminosity of broad H$\alpha$ emission line ($L_{\rm H\alpha^B}$).
The calculated $\log R$ data are recorded in Table~\ref{tab:R}.
The $R$ values of the 125 newly found sources
range from 0.9 to 181.3 with a median of 12.0,
while the $R$ of the 26 known sources
from 6.3 to 314.1 with a median of 29.6.

\begin{figure}[htb]
\centering
\includegraphics[width=0.48\textwidth]{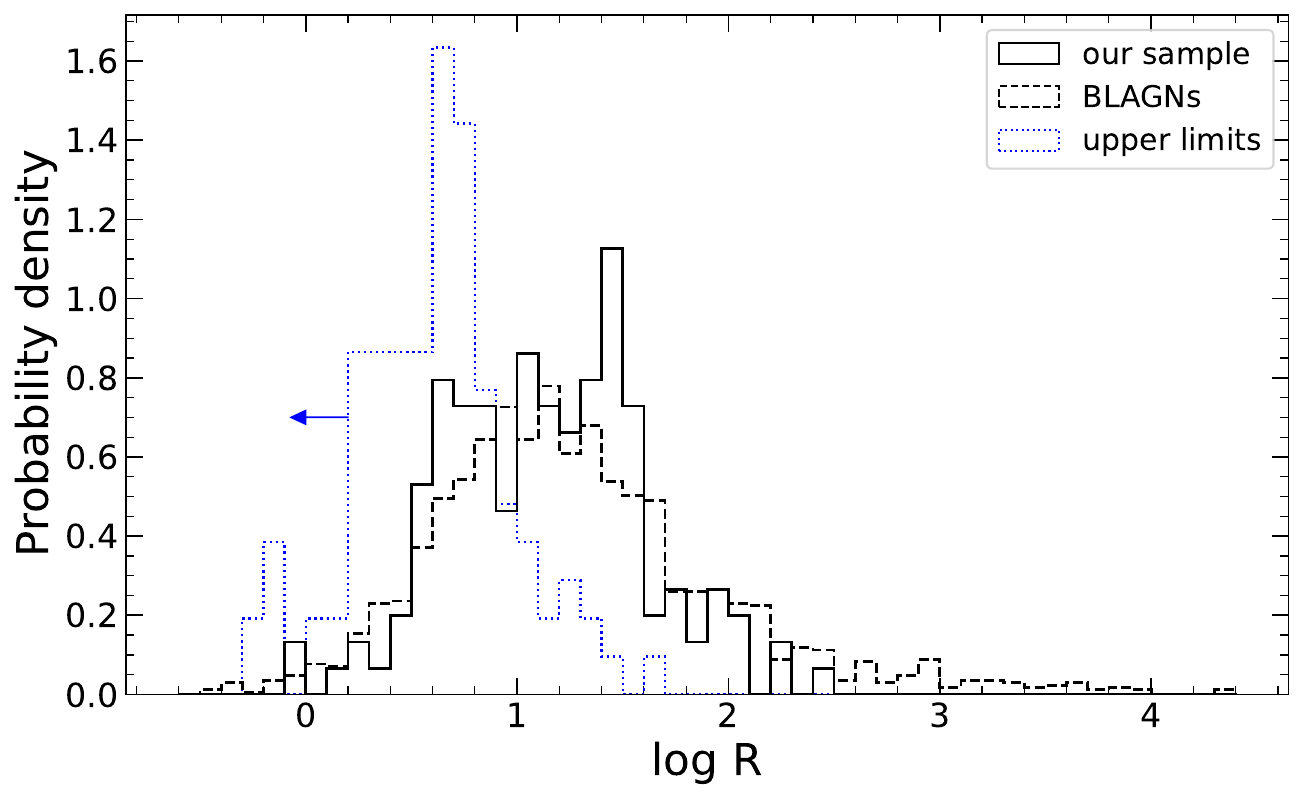}
\caption{Normalized histogram of the radio loudness of our radio sample (black solid line).
	The normalized histograms of the so-called BLAGNs (black dashed line)
	 and the upper-limit sources (blue dotted line) are also plotted in the same way as in Figure~\ref{fig:Lhist}. }
\label{fig:Rhist}
\end{figure}

For comparison, similar to the $P_{\rm 20cm}$ case,
we also calculate the radio loudness of the BLAGNs.
The normalized histograms of $\log R$ of our sample and the BLAGNs are plotted in Figure~\ref{fig:Rhist}.
Both distributions are similar.
But here a caveat is in order:
the distributions of both samples are only for the detected radio sources
from the surveys such as FIRST (and LoTSS for our sample);
if the upper-limit sources are taken into account, the distributions certainly would
be moved and skewed to small-$R$ end (cf. Figure~\ref{fig:Rhist}).

If the conventional radio-loud criterion $R>10$ is taken,
23 of the 26 known sources (\citealt{hyLiu2018imbh})
and 70 of our 125 brand-new sources are radio-loud.
Such a high radio-loud fraction may be indicative of considerable radio contributions from their host galaxies;
this is particularly possible considering the low optical luminosity of these low-mass AGNs.

\begin{figure*}[htb]
  \centering
  \includegraphics[width=1\textwidth]{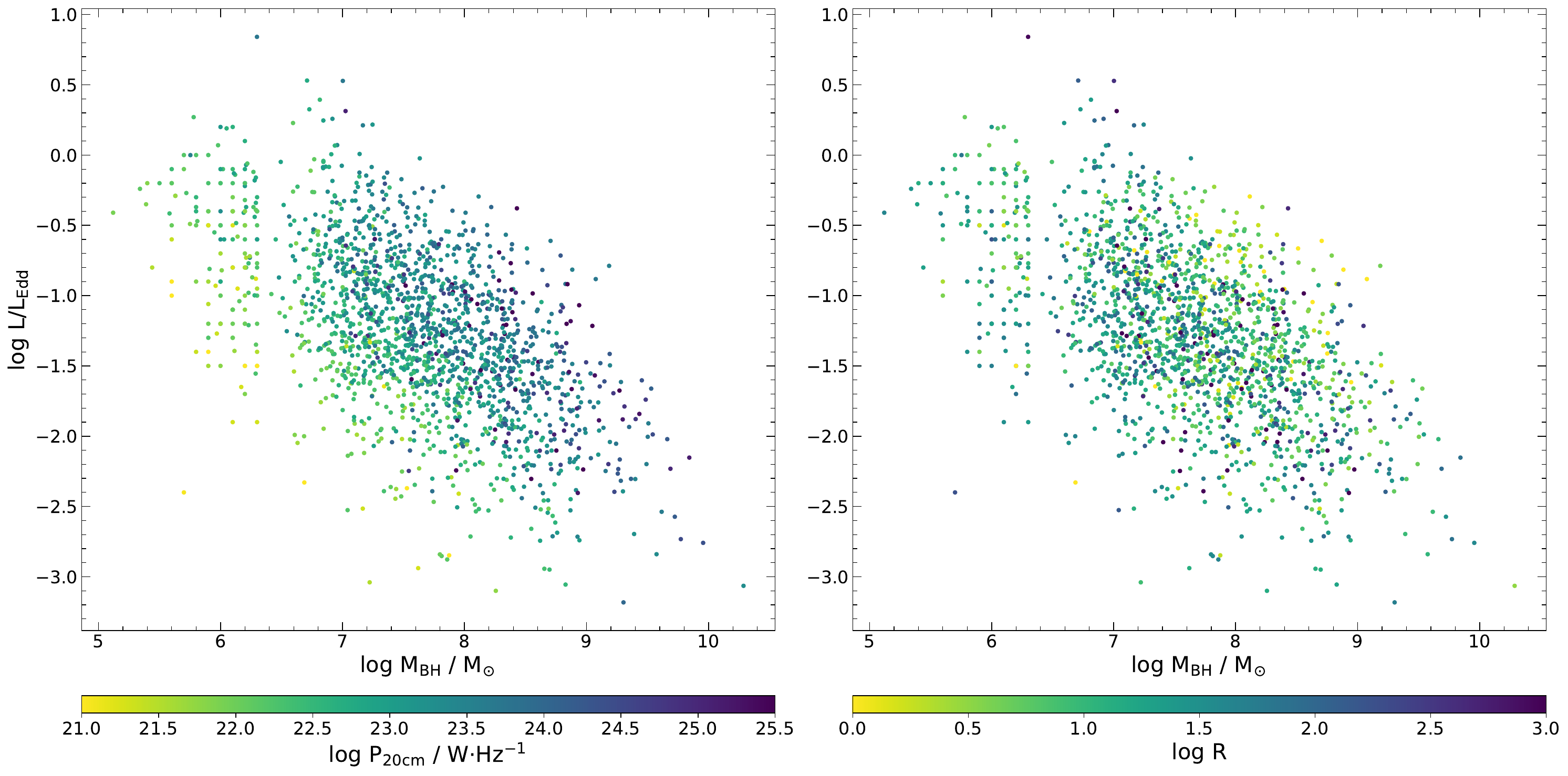}
  \caption{Distributions of the rest-frame power at 20cm ($P_{\rm 20cm}$; left panel) and radio loudness ($R$; right panel)
  	on the diagram of black hole mass versus Eddington ratio for the radio sample of 151 low-mass AGNs,
  	together with all the other radio sources in the SDSS-DR7 broad-line AGN sample (dubbed BLAGNs in the text).
  	The radio power and loudness are color-coded according to the respective colorbars.
    }
  \label{fig:power_R_on_M-lratio-diagram}
\end{figure*}

\subsection{Dependence or not of radio power and loudness on $\mbh$ and $\lratio$ in Seyferts}

A linear correlation in logarithmic space has been found
among BH mass $M_{\rm BH}$, and radio and X-ray powers,
which forms a so-called fundamental plane (FP) of BH activity 
\rev{\citep{2003MNRAS.345.1057M, 2004A&A...414..895F, 2006A&A...456..439K, 2009ApJ...706..404G,2014ApJ...788L..22G}}.
Similarly, the FP can be translated into the plane
of radio loudness with \mbh\ and the bolometric Eddington ratio \lratio\
(see, e.g., \citealt{xie2017, Qian+2018}).
Safely and strictly speaking,
the BH FP is restricted to AGNs with radiatively inefficient accretion flows
with $\lratio \lesssim 10^{-2}$.
For radiatively efficient AGNs such as Seyferts,
the existence or the concrete coefficients of such a BH FP are still controversial 
\rev{
\citep{2013IAUS..290...29H, 2006A&A...456..439K, 2011MNRAS.414..677C, 2014ApJ...787L..20D}.
} 

In light of the data quality of the present sample (e.g., concerning the accurate radio emission from the jets),
\rev{to avoid any possible confusion or even misleading}
here we do not attempt to model/fit the FP relationship in Seyferts or IMBH AGNs,
but conduct correlation and partial correlation tests instead, which are less demanding on data quality
than fitting a FP relation.

For this purpose, we use all the reliable radio data
for both our low-mass AGN sample
and the BLAGNs described in \S\ref{sect:Power} and \S\ref{sect:Loudness}.
Note that the BLAGNs denote the other broad-line AGNs
(namely, more massive Seyfert 1s) in the sample of \cite{hyLiu2019dr7All},
and that their radio data are retrieved from the official FIRST catalog.
The data of \mbh\ and \lratio\ are obtained from
the respective catalogs of \citep{Dong2012, hyLiu2018imbh, hyLiu2019dr7All}.

Viewing from the left panel of Figure~\ref{fig:power_R_on_M-lratio-diagram},
there are apparent correlations between any two quantities of
$\log P_{\rm 20cm}$, $\log \lratio$ and $\log M_{\rm BH}$.
The one between $\log \lratio$ and $\log M_{\rm BH}$,
is well known in any optically selected samples
and caused by selection effects and cosmic downsizing (see \citealt{Dong2012} for the detail).
The other two concern $P_{\rm 20cm}$ correlating positively with \mbh\ and \lratio, respectively.

In order to minimize the effects caused by inter-dependence among the three quantities
and to pin down the intrinsically significant correlations,
we perform various Spearman partial correlation tests to the above two correlations concerning radio power.
First, we conduct the following tests:
for the correlation between $P_{\rm 20cm}$ and \lratio\ controlling for \mbh,
$r_{\rm s}$ (Spearman correlation coefficient) is 0.48 and
$P_{\rm null}$ (probability for null hypothesis that there is no correlation) is $1.2 \times 10^{-104}$;
for that between $P_{\rm 20cm}$ and \mbh\  controlling for \lratio,
$r_{\rm s}=0.67$ and $P_{\rm null}=2.4 \times 10^{-245}$.
Besides, these two correlations may be affected by the dependence of the three variables on redshift in common.
Thus we further control the redshift in the above two partial correlation tests.
As a result,
\pnull\ becomes $1.1 \times 10^{-3}$ for that between $P_{\rm 20cm}$ and \lratio\ (controlling for both \mbh\ and redshift),
whereas $\rs = 0.26$ and $\pnull = 9.8 \times 10^{-29}$
for that between $P_{\rm 20cm}$ and \mbh\ (controlling for both \lratio\ and redshift).
Thus we conclude that
for Seyfert galaxies
radio power correlates intrinsically and positively with \mbh,
yet marginally at most with \lratio.

We also have test correlations among $\log R$, $\log \lratio$ and $\log M_{\rm BH}$.
There are no strong correlations of $\log R$ with either \mbh\ or \lratio.
This can be seen easily from
the right panel of Figure~\ref{fig:power_R_on_M-lratio-diagram},
in which $\log R$ nearly distributes uniformly
in the plane of $\log \lratio$ and $\log M_{\rm BH}$.
This may be due to large scatter in the $\log R$ data  compared with
its small dynamic range ($\log R$ mainly in the range between 0 and 2.5).
Thus we cannot say anything about the correlations concerning radio loudness.

Note that in Figure~\ref{fig:power_R_on_M-lratio-diagram}
the gap between low-mass AGNs and other Broad-line AGNs
are due to the difference in the viral mass formulae
used in the three catalog papers.
But the gap has little influences on the above Spearman tests
because Spearman method is rank ordered.
We are still working on an expanded sample of broad-line AGNs (W.-J.~Liu \etal\ 2023, in preparation),
in the present work we thus do not investigate further the radio properties of the whole Seyfert-1 population.

\begin{deluxetable*}{cccccccc}
  \tablewidth{0pt}
  \setlength{\tabcolsep}{8pt}
  \tabletypesize{\scriptsize}
  \tablecaption{Physical properties of the radio low-mass AGNs \label{tab:R}}
  \centering
  \tablehead{
    \colhead{ID} &\colhead{$z$} &\colhead{$\log M_{\rm BH}$} &\colhead{$\log L/L_{\rm Edd}$} &\colhead{$\log L_{\rm H\alpha^B}$} &\colhead{$\log L_{5100}$} &\colhead{$\log P_{\rm 20cm}$} &\colhead{$\log R$} \\
    \colhead{} &\colhead{} &\colhead{log(M$_\sun$)} &\colhead{} &\colhead{log(erg/s)} &\colhead{log(erg/s)} &\colhead{log(W/Hz)} &\colhead{}
    \\
    \colhead{(1)} &\colhead{(2)} &\colhead{(3)} &\colhead{(4)} &\colhead{(5)} &\colhead{(6)} &\colhead{(7)} &\colhead{(8)}
        }
  \startdata
   S1 & 0.0485 & 6.00 & -0.70 & 40.91 & 42.44 & 21.60 & 0.71\\
   S2 & 0.0733 & 6.20 & -1.00 & 40.77 & 42.31 & 21.77 & 1.00\\
   S3 & 0.0611 & 6.10 & -1.30 & 40.21 & 41.83 & 22.03 & 1.74\\
   F1 & 0.0687 & 6.00 & -0.90 & 40.73 & 42.28 & 22.31 & 1.57\\
   F2 & 0.0485 & 6.00 & -0.70 & 40.91 & 42.44 & 21.62 & 0.73\\
\enddata
  \tablecomments{Column (1), identification number
  	assigned in Tables~\ref{tab:sample_skaPs} and \ref{tab:sample_deepFIRST}.
  	Columns (2) to (5), the redshift, virial mass of black hole, Eddington ratio and luminosity of broad H$\alpha$ emission line,
  	retrieved from the two parent optical catalogs.
  	Column (6), the continuum luminosity at rest-frame 5100 \AA \; ($L_{5100}\equiv \lambda L_\lambda \;{\rm at}\; \lambda=5100\;$\AA), derived from $L_{\rm H\alpha^B}$.
  	Column (7), the calculated radio power at rest-frame 20 cm.
  	Column (8), radio loudness. \\
  (This table is available in its entirety in a machine-readable form in the online journal. A portion is shown here for guidance regarding its form and content.)}
\end{deluxetable*}


\subsection{The \logNlogS/ test and predictions \label{section:logNS}}

Motivated by the \logNlogS/ exploration in the FIRST-catalog paper (\citealt{White97-FIRSTCatalog}),
with great interest we assess the completeness and depth of our radio sample
by carrying out the similar exploration but down to a fainter limit ($S$).
In addition, with a certain intellectual curiosity
we predict the detectability of the entire 513 optical selected low-mass AGNs in the parent sample
in terms of the \logNlogS/ relation.

In a non-expanding spatially flat universe, the so-called Euclidean universe,
this relation  is $N \propto S^{-1.5}$
if the observed population has a constant space density,
where $N$ is the number of all the sources having flux above the limiting flux $S$
(see \S10.1 of \citealt{1997iagn.book.....P} for the detail).
Our Universe is spatially flat yet expanding,
and thus the \logNlogS/ relation is not so simple (actually without an analytic expression).
But once the cosmological parameters are given,
the relation is certain (see Appendix~\ref{appendix:logNlogS}).
Thus, if we know \textit{a priori} that the space density of the sample of interest
is constant,
we can use the \logNlogS/ relation to judge to what a limiting flux the sample is complete.

In this respect, we regard the radio emission of the whole galaxies
(Seyfert nuclei plus their host galaxies) to be from the same population.
The practical reason is that
in light of the spatial resolution we cannot separate the nuclei emission
from the host-galaxy one.
The second and positive reason is that
the radio properties of general Seyfert galaxies
(unlike quasars that can either have large-scale jets or not;
see \S\ref{sect:properties}; \citealt{2008ARA&A..46..475H}),
as well as their host-galaxy properties (see, e.g., \citealt{Dong2012}),
appear indeed belong to a same population.

\subsubsection{Distribution on the \logNlogS/ plane \label{sect:logNS_data}}

For our radio detected low-mass AGNs,
we cannot treat them as a single sample to make the \logNlogS/ test,
because they come from different radio surveys with different sensitivities and
different survey areas.
Below we use two groups of our radio sources, one from the FIRST survey
(including our deeply mined and the officially cataloged;
hereafter called FIRST sources for short)
and the other from LoTSS-DR2,
because both surveys have large areas and their sources dominate our radio sample.
The integrated flux densities (physically, nucleus $+$ host emission) are used.

\begin{figure*}[htb]
  \centering
  \includegraphics[width=0.9\textwidth]{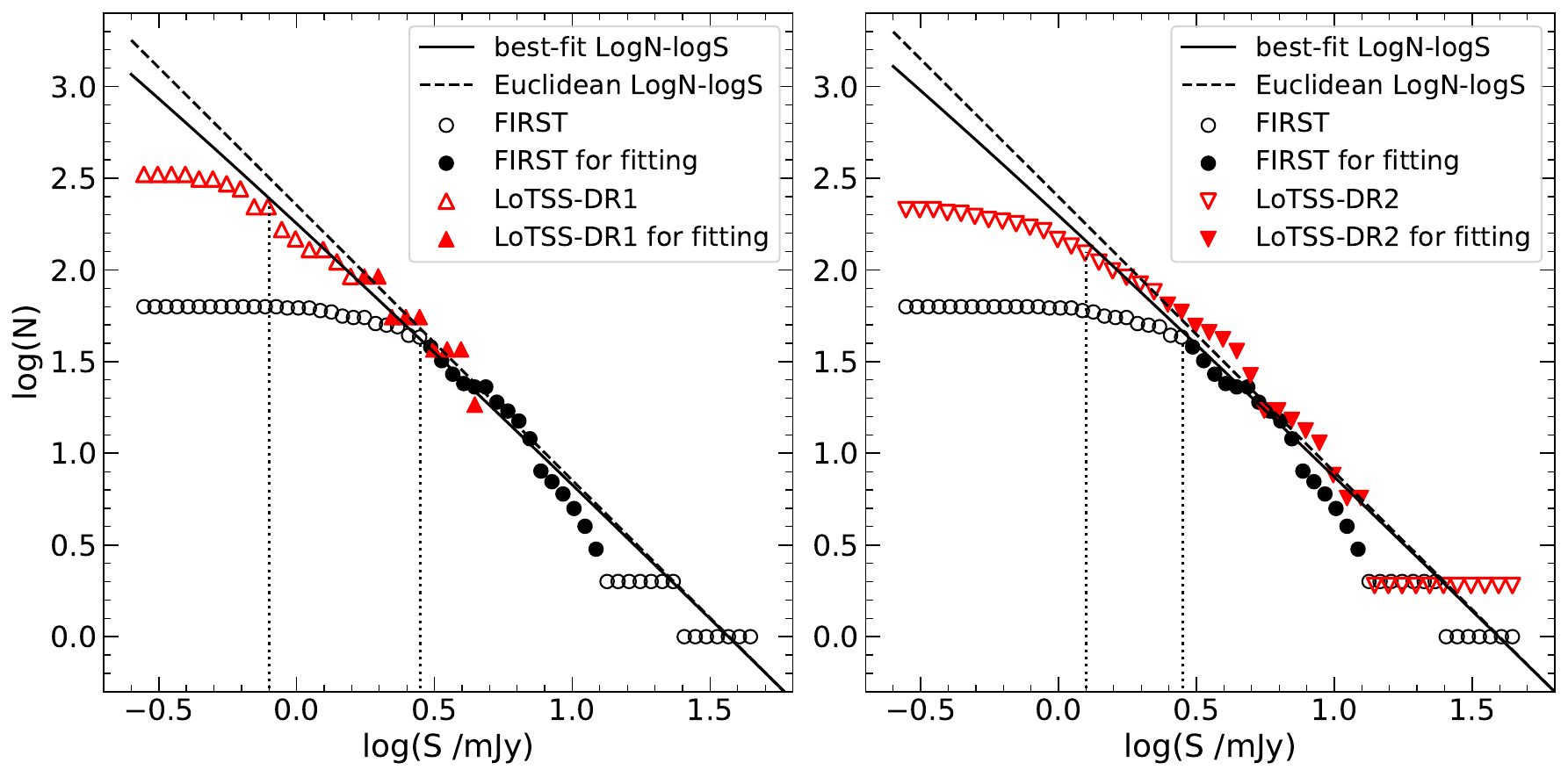}  
  \caption{The \logNlogS/ distributions of the LoTSS sources (DR1, red upward
  triangles on left panel; DR2, red downward triangles on right panel) 
  and FIRST sources (black circles on both panels).
  The flux densities are at 0.15 GHz;
  the source numbers, scaled in the LoTSS cases,
  correspond to the sky area of our FIRST sources (see \S\ref{sect:logNS_data}).
  The canonical \logNlogS/ relation for the static Euclidean universe
  ($N \propto S^{-1.5}$; dashed line in every panel)
  is plotted for comparative purpose.
  We also plot the \logNlogS/ model (solid curve;
  designed for the expanding, spatially flat Universe) that is fitted to
  the aforementioned distributions of our radio sample.
  \rev{Note that only the data points denoted with filled symbols are used for the fittings.}
  The dotted lines indicate the limiting flux densities of completeness for LoTSS and FIRST sources.
  }
  \label{fig:logNS_wData}
\end{figure*}

We plot the \logNlogS/ distributions of both FIRST and LoTSS-DR2 sources
in the same diagram (right panel of Figure~\ref{fig:logNS_wData}),
for the ease of checking the completeness of the two groups
and the overall behavior of our radio sample.
For putting them in the same figure,
the two groups need to have flux densities at the same frequency,
and have source numbers corresponding to an equal survey area.
First, we convert flux densities of FIRST sources at 1.4~GHz
to that at 0.15~GHz (the central frequency of LoTSS) adopting the spectral index of 0.46 used above.
Second, we scale the number of LoTSS-DR2 sources
to that corresponding to the survey area of FIRST,
by multiplying the ratio of their survey areas (\mbox{7813/4111}).
Here, 7813 (deg$^2$) is the overlapped survey area
of FIRST and the Spectroscopic Legacy of SDSS-DR7 (where our parent sample come);
4111 (deg$^2$) is the overlapped survey area of LoTSS-DR2
and SDSS-DR7 Spectroscopic Legacy.

We also plot the \logNlogS/ distributions of LoTSS-DR1 along with FIRST sources
(left panel of Figure~\ref{fig:logNS_wData});
this is used to verify the accuracy of
predicting source number with the \logNlogS/ relation (see \S\ref{sect:logNS_prediction}).

\subsubsection{Fitting the \logNlogS/ relation \label{sect:logNS_modeling}}

On the theoretical side, however,
the situation is a little complicated
because our real Universe is a spatially flat yet expanding universe,
not an Euclidean one.
In this Universe, the theoretical \logNlogS/ relation
deviates from the simple power law
(namely the canonical $N \propto S^{-1.5}$ relation of the Euclidean universe).
Here we give a brief description, and refer the reader to
Appendix~\ref{appendix:logNlogS} for the detail.
First of all, because our parent sample (the optically selected low-mass AGNs)
is limited to low redshifts $z <0.35$,
we can simply assume that our radio sources have a constant comoving density $n_0$.
Then the $N-S$ relation is expressed as follows,
\begin{equation}
  \begin{split}
  N(S) & =  \int\int_0^{d_\mathrm{co}(z_\mathrm{max})} n(r) r^2 \phi(L)\mathrm{d}r \mathrm{d}L \\
  & = \int \frac{n_0}{3} d^3_\mathrm{co}(z_\mathrm{max})\phi(L)\mathrm{d}L ~,
  \label{eq:NS_phiL}
  \end{split}
\end{equation}
where $\phi(L)\mathrm{d}L$ is the luminosity function;
$z_{\rm max}$ and $d_\mathrm{co}$ are described in Appendix~\ref{appendix:logNlogS}.
Note that $z_{\rm max}$, or equivalently $d_\mathrm{co}(z_{\rm max})$,
is a function of $S/L$ (refer to Equation~\ref{eq:S-zmax}),
not of $L$ explicitly;
i.e., the functional forms of $z_{\rm max}$ and $d_\mathrm{co}(z_{\rm max})$
can be constructed in a way that
they are not sensitive to the concrete values of luminosity
(see Appendix~\ref{appendix:logNlogS}).
Moreover, the distribution of the radio luminosities of our sample
is somehow narrow and well peaked (see Figure~\ref{fig:Lhist}).
Taking the above two points into account,
in order to derive a theoretical, expansion-modified \logNlogS/ relation
we thus simplify the luminosity distribution of
our sample with an effective luminosity (e.g., around the median luminosity).
Then Equation~\ref{eq:NS_phiL} is fully represented
by the system of Equations \ref{eq:S-zmax} and \ref{N-Dco(max)}.
As discussed in Appendix~\ref{appendix:logNlogS},
this equation system can be solved numerically only,
but we can model the numerical $N(S)$ relation with a carefully constructed
function of the modified double power-law form (Equation~\ref{eq:NS_doublePL}).
This model $N(S)$ curve is devised thus that 
the curve shape is controlled by three parameters ($\beta, \gamma$ and $d_\mathrm{t}$)
that are independent of the limiting flux $S$ and the luminosity $L$
(when $S$ and $L$ are normalized; see the text around
Equation \ref{eq:NS_doublePL} for the detail).

\rev{
	We would like to note that 
	the steep radio luminosity function of AGNs in the related luminosity range 
	\citep{2014ARA&A..52..589H, 2005A&A...435..521N, 2007MNRAS.375..931M, 2017ApJ...846...78Y, 2018A&A...616A.152S} 
	further supports the above simplification of a single effective luminosity.
    The intrinsic radio luminosity function of low-redshift AGNs
    in our luminosity range has a fairly steep slope, 
    $\alpha_\mathrm{LF} \approx$ \mbox{0.7--0.78},
    if a single power-law ($n \propto L^{-\alpha_\mathrm{LF}}$) is fitted to the data 
	\citep{2005A&A...435..521N, 2018A&A...616A.152S}.
By Estimating in terms of Equation~\ref{our_N-S},
we can sense that with decreasing $L$ the power-law increase of the $n_0$ term and the decrease of $L$ term in that equation
counterbalance each other.
We notice also that those studies reported an intrinsic turnover begins 
	at around $P_{\rm 20cm} \approx 10^{19.5}$ W Hz$^{-1}$ toward low-luminosity end
	of the radio luminosity function of low-redshift AGNs.
	But this turnover luminosity is six times smaller than 
	the smallest $P_{\rm 20cm}$ of our sample (see Figure~\ref{fig:Lhist}).
	As the above studies stated, 
	the radio luminosity function in the neighboring (higher) luminosity range 
	continues with the same slope 
	down to at least $P_{\rm 20cm} \approx 10^{20}$ W Hz$^{-1}$. 
	Thus, our present sample is immune to this low-luminosity turnover in the radio luminosity function.
} 

We now fit the model $N(S)$ function
to the distributions of our source counts as plotted in Figure~\ref{fig:logNS_wData}.
For the ease of narration, we put the model function here
(see also Equation~\ref{eq:NS_doublePL}),
\begin{equation}
  N(S)  =  C\left(\frac{S/S_0}{L^\prime/L_0}\right)^{-1.5}
  \left[d_t+\left(\frac{S/S_0}{L^\prime/L_0}\right)^{-\gamma}\right]^{\frac{1.5+\beta}{-\gamma}}~.
  \label{eq:NS_Lprime}
\end{equation}
Here the shape parameters $\beta, \gamma$ and $d_\mathrm{t}$ are fixed to
the theoretical values obtained in Appendix~\ref{appendix:logNlogS};
$\beta = -8.339\times10^{-3}$, $\gamma = 0.392$, and $d_t = 12.36$.
The normalization factors $S_0$ and $L_0$
are fixed to 1~mJy and $10^{22.57}$ W$\cdot {\rm Hz}^{-1}$
at 0.15 GHz
(the median luminosity of our data plotted in Figure~\ref{fig:logNS_wData}),
the same as adopted in Appendix~\ref{appendix:logNlogS} (cf. Figure~\ref{fig:Append-logNS-expanding}).
The effective luminosity, $L^\prime$,
is fixed to the median luminosity (being equal to $L_0$);
this brings little error considering that
the distribution of the radio luminosities of our sample is narrow and fairly peaked,
and that model $N(S)$ is not sensitive to the value of $L^\prime$
(note that $L^\prime$ just shifts the $N(S)$ curve horizontally).
$C$, mainly related to the unknown quantity $n_0$ and responsible for shifting the
model $N(S)$ vertically, is a free parameter
and obtained by fitting the model $N(S)$ to the \logNlogS/
distributions of our sources.

A technical detail should be described, concerning the fitting range of the data.
Obviously the low-flux ends of both FIRST and LoTSS data are not complete,
and thus should not be included in the fitting.
We determine the fitting range in this way:
first we get the highest redshift of an observed data set (either FIRST or LoTSS),
treat it as the maximum redshift $z_\mathrm{max}$ (corresponding to the limit flux $S$) of
the theoretical \logNlogS/ curve (see Figure~\ref{fig:Append-logNS-expanding}),
and calculate the slope of \logNlogS/ at this $z_\mathrm{max}$ from the theoretical curve,
which is regarded to be the flattest of the possible pointwise slopes of the complete source counts range
(called `the possible flattest slope' for short);
then we fit our \logNlogS/ data with a modified double power law model and
calculate the slopes of this model at fluxes starting from the low-flux end upwards,
and find at which flux level the calculated slope gets steeper than `the possible flattest slope';
we take this flux level as the lower bound of the fitting range;
at the high flux end,
we only discard the points with so small source counts that large
fluctuations appear (these points distribute unnaturally horizontally).
As for the LoTSS-DR2 data set, the highest redshift is 0.33,
`the possible flattest slope' is $-1.262$,
and thus the fitting range is 2.49~mJy $\leqslant S \leqslant $ 12.48~mJy;
likewise, the fitting range of the FIRST data set is 3.06~mJy $\leqslant S \leqslant $ 12.20~mJy
(\rev{corresponding to the filled symbols in the right panel of Figure~\ref{fig:logNS_wData},
while the open symbols are not used for the fitting}).
In the fitting, the two data sets are used jointly.

The best-fit \logNlogS/ curve to our data set is plotted in right panel of Figure~\ref{fig:logNS_wData},
with best-fit $C=3.83\times10^6$.
The canonical Euclidean $\log N \propto -1.5\log S$ relation is also plotted as comparison.

\subsubsection{Assessing the completeness and depth \label{sect:logNS_completeness_test}}

With the best-fit \logNlogS/ curve in hand,
we can assess the completeness of our radio sample,  
following the approach of \citet[][see their \S6]{White97-FIRSTCatalog}.
\rev{
Just as we noted in \S\ref{sect:logNS_modeling},
the \logNlogS/ distribution of the present sample, including its turnover (getting flatter) toward lower-flux end,
is not affected by the low-luminosity turnover of radio luminosity function.
} 
From the right panel of Figure~\ref{fig:logNS_wData},
the distribution of the sources from the FIRST deep mining
agrees well with the best-fit \logNlogS/ relation downwards till
flux density of 2.79~mJy at 0.15~GHz (corresponding to \mbox{1 mJy} at 1.4~GHz),%
\footnote{There is significant deviation at high flux densities ($\gtrsim 10$ mJy),
which is because the number of the high-flux sources is
small, resulting in the large fluctuation.}
and gets flatter than the relation below this flux density,
which indicates that
the limiting flux density for the source completeness of our FIRST deep mining is 1~mJy at 1.4~GHz.
The thus-determined limiting flux density is half of that of the FIRST official catalog
(\mbox{$\approx$\,2} mJy at 1.4~GHz; see \S6 of \citealt{White97-FIRSTCatalog}).
Note that 
the limiting flux density of completeness
is not the same concept as the detection limit defined in the FIRST official catalog (1~mJy)
that is 5 times the rms noise plus the CLEAN bias (0.25~mJy).
Besides, the limiting flux density of completeness concerns the integrated flux density of a source,
whereas the detection limit is generally applied to the peak flux density
(see, e.g., the definition of the official FIRST catalog; \S4 of \citealt{White97-FIRSTCatalog});
the integrated and peak flux densities of a source are not the same, when the source is resolved.
Likewise, the limiting flux density for the completeness of our LoTSS-DR2 sources
turns out to be 1.26 mJy at 0.15\,GHz (namely, \mbox{0.45\,mJy} at 1.4\,GHz),
albeit conservatively determined
which is broadly consistent with the LoTSS-DR2 official estimation,
1.1~mJy for the 95\% completeness
(see \S3.6 of \citealt{2022A&A...659A...1S}).
Again, note that this completeness limit should not be confused with
the 5$\sigma$ detection limit (0.415~mJy).
The nominal detection limit concerns peak flux density,
and is just a median because the rms noise varies across the LoTSS-DR2 sky region
(see \S3.6 of \citealt{2022A&A...659A...1S} for the detail).

\subsubsection{The predictions \label{sect:logNS_prediction}}

Based on the best-fit \logNlogS/ curve, we can also make some predictions about
the respective numbers of radio detections of our parent sample
corresponding to various detection limits (as well as specific survey areas).
Certainly for this purpose
there is a premise that
as far as the total radio emission is concerned
these Seyfert 1 galaxies can be regarded as a same population;
as stated in the beginning of this Subsection (\S\ref{section:logNS}),
this is a reasonable assumption.

First, we predict how many the detected LOFAR counterparts will be once LoTSS is completed.
This can be made by keeping the current limiting flux density of completeness of the LoTSS-DR2 sources
(1.26 mJy; see the above) and simply expanding the survey area.
That is, multiplying the source number on the best-fit \logNlogS/ curve
at the limiting flux density of 1.26 mJy
 (see the vertical axis of the right panel of Figure~\ref{fig:logNS_wData})
by a factor of the area ratio (\mbox{8032/7813}).
Here 8032 deg$^2$ is the total area of Spectroscopic Legacy of SDSS-DR7,
which our parent sample covers;
the number 7813 is explained in \S\ref{sect:logNS_data}.
As a result, we predict that once LoTSS is completed
about 149 radio counterparts of our parent sample will be detected.

We have tested and verified this prediction in the following way:
using the sources from LoTSS-DR1 (cf. Table~\ref{tab:sample_skaPs})
to ``predict'' the number of the sources from LoTSS-DR2.
For this purpose,
we fit the \logNlogS/ model to the data set of LoTSS-DR1 and FIRST sources
(see the last paragraph of \S\ref{sect:logNS_data},
and the left panel of Figure~\ref{fig:logNS_wData}),
and find that
the limiting flux density of completeness for LoTSS-DR1 is 0.79 mJy at 0.15 GHz
(cf. \S\ref{sect:logNS_completeness_test}).
We now scale vertically this best-fit curve
by the area ratio \mbox{4111/7813}
to correspond to the actual sky area of LoTSS-DR2
 (see \S\ref{sect:logNS_data} for the two numbers and the detail of Figure~\ref{fig:logNS_wData}).
Reading from this scaled curve (at the limiting flux density of completeness 0.79 mJy),
we can ``predict'' (i.e., based on LOTSS-DR1)
that we can find 130 counterparts in LoTSS-DR2.
This ``prediction'' agrees reasonably with the actual number of LoTSS-DR2 sources (112),
considering a worse (and not homogeneous) completeness of LoTSS-DR2 than LoTSS-DR1
(see \citealt{2022A&A...659A...1S,2019A&A...622A...1S}).
If we use the nominal limiting flux density of completeness
for LoTSS-DR2 determined in \S\ref{sect:logNS_completeness_test},
i.e. ``predicting'' based on LoTSS-DR1 \textit{a posteriori},
then the predicted number of LoTSS-DR2 counterparts would be 68,
almost the same as the actual number of
LoTSS-DR2 counterparts with flux density $\geqslant$ 1.26 mJy (64 sources).

The second prediction is about the question:
to what a limiting flux density of completeness can
we find (in the statistical sense) all radio counterparts of the parent sample?
First, we scale vertically the \logNlogS/ curve presented in \S\ref{sect:logNS_data}
(this curve corresponds to a sky area of 7813 deg$^2$),
to match the total sky area of 8032 deg$^2$.
Then, along the curve, a limiting flux density is found
corresponding to the source number of 513,
which is 0.5 mJy at 0.15 GHz. 
\rev{
Here we should note that
the actual limiting flux density may turn out to be somehow lower;
this is because our parent sample being optically selected Seyfert 1s,
just like normal optical surveys being magnitude-limited,
is not complete at high-redshift end.
In addition, just as stated in \S\ref{sect:logNS_modeling},
the low-luminosity radio sources with 
$P_{\rm 20cm} \lesssim 10^{19.5}$ W Hz$^{-1}$
may have a smaller space density than the power-law extrapolation of 
the luminosity function of their neighbors not too luminous away.
If so, and if the radio powers of some yet-to-detect sources in our parent sample 
are below the turnover luminosity, 
then the actual limit of completeness should be further lower. }
\newline

\section{Summary \label{sect:summary}}
In this work, we have built a radio sample of low-mass AGNs,
by searching in large-scale radio continuum surveys
for the radio counterparts of 513 optically selected low-mass AGNs
(\citealt{Dong2012,hyLiu2018imbh}; the parent sample).
By exploiting the recent data releases of the surveys of SKA pathfinders,
particularly the second data release of LOFAR Two-metre Sky Survey (LoTSS DR2),
we have compiled a catalog of 117 radio low-mass AGNs with $5\sigma$ detection
(called the SKAPs subsample; Table~\ref{tab:sample_skaPs}),
which constitute the main body of our radio sample.
We have also deeply mined the FIRST images
with an elaborate procedure specialized for fitting faint radio sources,
and compiled a catalog of 63 radio low-mass AGNs
with $\mathrm{S/N} \geqslant 3.5$ (called deep FIRST subsample; Table~\ref{tab:sample_deepFIRST}).

Combining the two catalogs,
our radio sample comprises 151 distinct radio low-mass AGNs,
including 102 new reliable sources (S/N $\geqslant 5$) and
23 new candidates ($3.5 \leqslant \mathrm{S/N} <5$),
as well as 26 known sources (see \citealt{hyLiu2018imbh}).
Thus our sample is four times the record of radio low-mass AGNs (totaling about 40 sources previously).
Moreover, the majority of our newly found radio sources (119 of 125)
are fainter than the flux threshold of the official FIRST catalog (1 mJy at 1.4 GHz),
extending deeper into the low-flux regime
than those previous small samples of radio low-mass AGNs.

We have investigated the radio properties.
Only a small fraction (19/151) of the radio sources
exhibit ``extended'' morphology;
in particular, \object[SDSS J134738.24+474301.9]{J134738.24$+$474301.9}
appears to have a jet on tens of kpc scale.
Regarding the 125 newly found sources, their $P_{\rm 20cm}$ ranges
from $1.98 \times 10^{20}$ to $1.29 \times 10^{23}\;{\rm W\; Hz^{-1}}$
with a median of $1.12 \times 10^{22}\;{\rm W\; Hz^{-1}}$,
and their $R$ ranges
from 0.9 to 181.3 with a median of 12.0.
By combining the radio low-mass AGNs and
the radio counterparts in the FIRST catalog of a larger sample of broad-line AGNs
(i.e., radio Seyfert 1s at $z < 0.35$),
we have carried out partial correlation tests among
$P_{\rm 20cm}$ (or $R$), $\mbh$ and $\lratio$,
and found that for low-$z$ Seyfert 1s
$P_{\rm 20cm}$ correlates intrinsically and positively with $\mbh$,
yet only marginally with $\lratio$.

We have investigated the depth and completeness of our radio sample
(\S\ref{sect:logNS_completeness_test}),
in terms of the \logNlogS/ relation for the expanding, spatially flat Universe.
For this purpose, we carefully designed a modified double power-law function
as the \logNlogS/ model,
and fit it to the distributions of our LoTSS and FIRST sources.
The \emph{limiting flux densities of completeness} of the deeply mined FIRST sources
is pushed down to half the value of the FIRST official catalog
(2\,mJy; note the difference between the limit of completeness and the detection limit,
and see \S6 of \citealt{White97-FIRSTCatalog}).
The limiting flux density for the completeness of our LoTSS-DR2 sources
turns out to be 1.26 mJy at 0.15\,GHz (namely, \mbox{0.45\,mJy} at 1.4\,GHz).
Thus our radio sample is complete to
such a flux-density level that is deeper than \citet{White97-FIRSTCatalog} by a factor of 4.

We also predict in terms of the \logNlogS/ relation that
in total 149 radio counterparts of the parent sample will be detected once LoTSS is completed.
Further, when the limiting flux density of completeness is achieved to 0.5 mJy at 0.15 GHz,
all radio counterparts of the 513 optically selected low-mass AGNs in our parent sample
would be detected.

The present work kicks off our massive search for radio low-mass AGNs in the surveys of SKA pathfinders.
More and deeper data are to be released in near future, probably even from SKA-1.
We believe that it will be feasible to construct a sample of low-mass AGNs,
well defined and uniformly selected both in the optical and in the radio.
Besides, it will be valuable to search for new (and fainter) low-mass AGNs directly in the radio.

\section*{Acknowledgments}

This work is supported by National SKA Program of China (Nos. 2020SKA0110100, 2020SKA0120100,  
2020SKA0110102 and 2020SKA0110200),
Natural Science Foundation of China Grants
(NSFC 12373013, 
12203039, 
12373017, 
12173053, 
12192220, 12192223, 
12003047, 
and 11873083).
L.Q. and F.G.X. are also supported by the Youth Innovation Promotion Association of CAS 
(id. 2018075, no. Y2022027; no. Y202064). 
L.Q. is also supported by the CAS ``Light of West China'' Program.
J.-Z.~W. thanks Zhiyu Zhang and Junzhi Wang for the teaching of radio techniques.
X.-B.~D. thanks Luis Ho for the comments and discussions.

LOFAR data products were provided by the LOFAR Surveys Key Science project (LSKSP; https://lofar-surveys.org/) and were derived from observations with the International LOFAR Telescope (ILT). LOFAR (van Haarlem et al. 2013) is the Low Frequency Array designed and constructed by ASTRON. It has observing, data processing, and data storage facilities in several countries, which are owned by various parties (each with their own funding sources), and which are collectively operated by the ILT foundation under a joint scientific policy. The efforts of the LSKSP have benefited from funding from the European Research Council, NOVA, NWO, CNRS-INSU, the SURF Co-operative, the UK Science and Technology Funding Council and the J\"{u}lich Supercomputing Centre.

This work makes use of data from the Apertif system installed at the Westerbork Synthesis Radio Telescope owned by ASTRON. ASTRON, the Netherlands Institute for Radio Astronomy, is an institute of the Dutch Research Council ("De Nederlandse Organisatie voor Wetenschappelijk Onderzoek, NWO).

\appendix

\section{Other radio data in the literature}
\renewcommand{\thetable}{A\arabic{table}}
\setcounter{table}{0}

We conducted a search for other radio data, particularly by pointed observations,
for the low-mass AGNs in the parent sample (see Footnote~\ref{ftn:aboutA1A2}).
We find data for 10 sources, which are summarized in Table~\ref{tab:A1_otherData} below.
The four sources mentioned in \S\ref{sect:Cat_deep_FIRST},
which are in both deep FIRST and SKAPs subsamples,
are denoted with `S' in the ``Ref'' column.

\begin{deluxetable}{ccccccccc}[htb]
	\tablewidth{0pt}
	\tabletypesize{\scriptsize}
	\tablecaption{Other radio data for some sources in our sample \label{tab:A1_otherData}}
	\centering
	\tablehead{
		\colhead{ID} &\colhead{SDSS Name} &\colhead{$F_{\rm p}$} &\colhead{$F_{\rm p\_err}$} &\colhead{$F_{\rm i}$} &\colhead{$F_{\rm i\_err}$} &\colhead{Freq} &\colhead{SpecIndex} &\colhead{Ref}\\
		\colhead{} &\colhead{} &\colhead{mJy/beam} &\colhead{mJy/beam} &\colhead{mJy} &\colhead{mJy} &\colhead{GHz} &\colhead{} &\colhead{}\\
		\colhead{(1)} &\colhead{(2)} &\colhead{(3)} &\colhead{(4)} &\colhead{(5)} &\colhead{(6)} &\colhead{(7)} &\colhead{(8)} &\colhead{(9)}
	}
	\startdata
	F13 & J091449.06$+$085321.1 & \nodata & \nodata & 0.6 & 0.03 & 8.5 & 0.7 & (2) \\
	F25 & J110258.74$+$463811.5 & 5.52192 & 1.16947 & 6.66469 & 1.47344 & 0.144 & 0.7 & (1);S \\
	F36 & J121629.13$+$601823.5 & \nodata & \nodata & 0.37 & 0.03 & 8.5 & 0.7 & (2)  \\
    F38 & J124035.82$-$002919.5 & \nodata & \nodata & 0.45 & 0.03 & 8.5 & 0.7 & (2) \\
    & & 0.54 & 0.05 & 0.7 & 0.1 & 4.86 & 0.62 & (3) \\
    & & 1.8 & 0.09 & 1.88 & 0.13 & 1.4 & 0.76 & (4) \\
    & & 0.75 & 0.05 & \nodata & \nodata & 4.86 & 0.76 & (4) \\
    & & 0.44 & 0.03 & \nodata & \nodata & 8.46 & 0.76 & (4) \\
	F42 & J131957.07$+$523533.8 & 5.12 & 1.13401 & 6.91794 & 1.58265 & 0.144 & 0.7 & (1)\\
	F43 & J132041.00$+$283820.7 & \nodata & \nodata & 2.638 & 0.532 & 0.15 & 0.755 & (7);S \\
	F44 & J132428.24$+$044629.6 & \nodata & \nodata & 0.5 & 0.03 & 8.5 & 0.7 & (2) \\
	F47 & J134738.24$+$474301.9 & 3.81048 & 0.92878 & 6.53067 & 1.52429 & 0.144 & 0.7 & (1);S \\
	F51 & J144108.70$+$351958.8 & 1.78 & 0.23 & 3.56 & 0.32 & 0.15 & 0.79 & (6);S \\
	F54 & J152205.41$+$393441.2 & 1.03 & 0.021 & 2.03 & 0.16 & 1.6 & \nodata & (5) \\
	& & 0.329 & 0.013 & 0.38 & 0.039 & 5.2 & \nodata & (5) \\
	& & 0.202 & 0.008 & 0.273 & 0.021 & 9 & \nodata & (5) \\
	\enddata
	\tablerefs{(1) \cite{2017AA....598A.104S};
		(2) \citet{2014ApJ...788L..22G};
		(3) \citet{2006ApJ...636...56G};
		(4) \citet{2008ApJ...686..838W};
		(5) \citet{2020AA....636A..64B};
		(6) \citet{2016MNRAS.460.2385W};
		(7) \citet{2016MNRAS.462.1910H}.}
	\tablecomments{Column (1), identification numbers assigned in this paper.
		Column (2), SDSS name.
		Columns (3) to (6), peak and integrated flux densities, and their errors.
		Column (7), central frequency.
		Column (8), spectral index $\alpha _r$ (assuming $f_\nu \propto \nu^{-\alpha _r }$).
		Column (9), references for the radio data. J124035.82$-$002919.5 is in fact GH 10.
		J152205.41$+$393441.2 has multiple spectral indices; see \mbox{Ref.\,(5)} for the detail.
		}
\end{deluxetable}

\clearpage

As introduced in \S\ref{intro},
there are 40 reliable radio sources of low-mass AGNs prior to the present work,
of which 26 are included in our radio sample (see Footnote~\ref{ftn:aboutA1A2}).
For the other 14 sources,
we summarized their information from the literature in Table~\ref{tab:A2_otherSources} below.

\begin{deluxetable}{cccccccc}[htb]
  \tablewidth{0pt}
  \tabletypesize{\scriptsize}
  \tablecaption{Other radio sources in the literature \label{tab:A2_otherSources}}
  \centering
  \tablehead{
    \colhead{SDSS Name} &\colhead{$F_{\rm p}$} &\colhead{$F_{\rm p\_err}$} &\colhead{$F_{\rm i}$} &\colhead{$F_{\rm i\_err}$} &\colhead{Freq} &\colhead{SpecIndex} &\colhead{Ref}\\
    \colhead{} &\colhead{mJy/beam} &\colhead{mJy/beam} &\colhead{mJy} &\colhead{mJy} &\colhead{GHz} &\colhead{} &\colhead{}\\
    \colhead{(1)} &\colhead{(2)} &\colhead{(3)} &\colhead{(4)} &\colhead{(5)} &\colhead{(6)} &\colhead{(7)} &\colhead{(8)}
        }
  \startdata
  J082443.28$+$295923.5 & \rev{2.35} & \rev{0.14} & 1.77 & 0.17 & 1.4 & \nodata & (1,7) \\
                        & \nodata & \nodata & 0.93 & 0.05 & 8.5 & 0.7     & (2) \\
  J101246.49$+$061604.7 & \rev{1.57} & \rev{0.22} & 1.03 & 0.23 & 1.4 & \nodata & (1,7) \\
                        & \nodata & \nodata & 0.42 & 0.03 & 8.5 & 0.7     & (2) \\
  J105108.81$+$605957.2 & \rev{2.75} & \rev{0.14} & 2.41 & 0.18 & 1.4 & \nodata & (1,7) \\
                        & \nodata & \nodata & 0.54 & 0.03 & 8.5 & 0.7     & (2) \\
  J110501.97$+$594103.6 & \rev{4.83} & \rev{0.17} & 5.96 & 0.34 & 1.4 & \nodata & (1,7) \\
                        & \nodata & \nodata & 1.10 & 0.06 & 8.5 & 0.7     & (2) \\
  J131659.37$+$035319.8 & \rev{2.00} & \rev{0.15} & 1.55 & 0.17 & 1.4 & \nodata & (1,7) \\
                        & \nodata & \nodata & 0.57 & 0.03 & 8.5 & 0.7     & (2) \\
  J155909.62$+$350147.4 & \rev{3.55} & \rev{0.13} & 3.39 & 0.22 & 1.4 & \nodata & (1,7) \\
                        & \nodata & \nodata & 0.58 & 0.04 & 8.5 & 0.7     & (2) \\
  J140829.26$+$562823.5 & \rev{1.30} & \rev{0.15} & 1.91 & 0.18 & 1.4 & \nodata & (1,7) \\
  J084029.91$+$470710.4 & 0.101 & 0.002 & 0.142 & 0.005 & 6 & \nodata & (3) \\
  J090613.76$+$561015.1 & 1.451 & 0.005 & 1.444 & 0.009 & 6 & \nodata & (3) \\
  J095418.15$+$471725.1 & 0.027 & 0.002 & 0.035 & 0.005 & 6 & \nodata & (3) \\
  J144012.70$+$024743.5 & 0.546 & 0.010 & 0.540 & 0.018 & 6 & \nodata & (3) \\
  J152637.36$+$065941.6 & 0.015 & 0.002 & 0.023 & 0.005 & 6 & \nodata & (3) \\
  NGC 4395              & 1.54 & \nodata & 1.68 & \nodata & 1.4 & 0.6 & (4) \\
                        & 0.68 & \nodata & 0.80 & \nodata & 5 & 0.6 & (4) \\
  NGC 404               & 2.10 & 0.08 & 2.83 & 0.14 & 1.5 & 1.08 & (5,6) \\
                        & 0.25 & 0.01 & 0.66 & 0.06 & 5 & 1.08 & (5,6) \\
                        & 0.11 & 0.01 & 0.47 & 0.04 & 7.5 & 1.08 & (5,6) \\
                        & 0.04 & 0.01 & 0.29 & 0.03 & 15 & 1.08 & (6) \\
  \enddata
  \tablerefs{(1) \cite{GH07sample};
(2) \citet{2014ApJ...788L..22G};
(3) \citet{2022MNRAS.516.6123G};
(4) \citet{2001ApJS..133...77H};
(5) \citet{2012ApJ...753..103N};
(6) \citet{2017ApJ...845...50N};
\rev{(7) FIRST official catalog.}}
  \tablecomments{Definitions of the columns are the same as in Table \ref{tab:A1_otherData}.
  	Some of these sources had VLBI observations:
  	J090613.76$+$561015.1 \citep{2020MNRAS.495L..71Y},
  	NGC 4395 \citep{2001ApJ...553L..23W, 2006ApJ...646L..95W, 2022MNRAS.514.6215Y},
  	NGC 404 \citep{2014ApJ...791....2P},
  	J082443.28$+$295923.5, J110501.97$+$594103.6 and J131659.37$+$035319.8 \citep{2022ApJ...941...43Y}.}
\end{deluxetable}

\clearpage

\section{Probable radio sources with $3\leqslant{\rm S/N}<3.5$ from VLA-FIRST images \label{appendix:first_3_3.5}}

Below we list the 23 probable radio counterparts with $3\leqslant{\rm S/N}<3.5$
of the optically selected broad-line AGNs,
which are deeply mined from the FIRST images as described in \S\ref{sect:DeepMiningFIRST}.
Because there exist optical AGNs at their well-specified sky positions,
according to Bayesian argument the genuineness of the radio emission
is more plausible than traditional blind-source detections with the same S/N (see also \citealt{White2007_noisy}).
Anyway, as optimal, probable candidates of radio low-mass AGNs, these sources deserve
deep and high-resolution observations in the future.

\renewcommand{\thetable}{\Alph{section}\arabic{table}}
\setcounter{table}{0}

\begin{deluxetable}{ccccccccccccc}[htb]
  \tablewidth{0pt}
  \tabletypesize{\scriptsize}
  \tablecaption{Low-mass AGNs deeply mined from the FIRST images with $3\leqslant{\rm S/N}<3.5$ \label{tab:snr_gt3}}
  \centering
  \tablehead{
    \colhead{ID} &\colhead{SDSS Name} &\colhead{P} &\colhead{$z$} &\colhead{RA} &\colhead{DEC} &\colhead{$F_{\rm p}$} &\colhead{$F_{\rm i}$} &\colhead{$F_{\rm i\_tot}$} &\colhead{rms} &\colhead{PA} &\colhead{MajAx} &\colhead{MinAx}\\
    \colhead{} &\colhead{} &\colhead{} &\colhead{} &\colhead{deg} &\colhead{deg} &\colhead{mJy/beam} &\colhead{mJy} &\colhead{mJy} &\colhead{mJy/beam} &\colhead{deg} &\colhead{arcsec} &\colhead{arcsec}\\
    \colhead{(1)} &\colhead{(2)} &\colhead{(3)} &\colhead{(4)} &\colhead{(5)} &\colhead{(6)} &\colhead{(7)} &\colhead{(8)} &\colhead{(9)} &\colhead{(10)} &\colhead{(11)} &\colhead{(12)} &\colhead{(13)}
        }
  \startdata
  F64.1 & J030417.78+002827.4 & 1 & 0.0445 & 46.07384 & 0.47450 & 0.555 & 1.364 & 2.852 & 0.104 & 113.4 & 10.67 & 7.96\\
  F64.2 & J030417.78+002827.4 & 1 & 0.0445 & 46.07425 & 0.47243 & 0.563 & 0.814 & 2.852 & 0.104 & 151.1 & 7.85 & 6.36\\
  F64.3 & J030417.78+002827.4 & 1 & 0.0445 & 46.07564 & 0.47141 & 0.576 & 0.674 & 2.852 & 0.104 & 148.1 & 9.09 & 4.45\\
  F65.1 & J073631.83+383058.4 & 1 & 0.0733 & 114.13377 & 38.51710 & 0.643 & 1.120 & 1.627 & 0.136 & 50.4 & 11.14 & 4.56\\
  F65.2 & J073631.83+383058.4 & 1 & 0.0733 & 114.13525 & 38.51852 & 0.663 & 0.508 & 1.627 & 0.136 & 9.9 & 4.80 & 4.65\\
  F66 & J083211.57+502408.5 & 1 & 0.1719 & 128.04943 & 50.40178 & 0.733 & 0.624 & 0.624 & 0.145 & 40.4 & 5.51 & 4.50\\
  \enddata
  \tablecomments{Columns (1) to (3) and (5) to (13), see Table~\ref{tab:sample_deepFIRST} for the definitions.
  Column (10) is the same as the $F_{\rm p\_e}$ column in Table~\ref{tab:sample_deepFIRST}, both being the local rms noise of the images.
  Column (4), redshift; retrieved from the parent optical catalogs.
  Note that J152637.37$+$065941.7 has been detected by deep VLA observation \citep{2022MNRAS.516.6123G}.\\
  (This table is available in its entirety in a machine-readable form in the online journal. A portion is shown here for guidance regarding its form and content.)}
\end{deluxetable}


\section{Low-mass AGNs with radio upper limits from LoTSS-DR2 \label{appendix:LoTSS-upperlimits}}

LoTSS-DR2 is both deep and wide, covering now about half of the sky area of our parent sample.
There are 104 low-mass AGNs in the parent sample that are covered by LoTSS-DR2 yet under detection.
Here we provide the information of those upper-limit sources.
The upper-limit flux density is set to be 5 times the local rms of each source,
where rms is retrieved from the official rms map of LoTSS-DR2.

\renewcommand{\thetable}{\Alph{section}\arabic{table}}
\setcounter{table}{0}

\begin{deluxetable}{ccccccccccc}[htb]
  \tablewidth{0pt}
  \tabletypesize{\scriptsize}
  \tablecaption{Low-mass AGNs in the parent sample with 5$\sigma$ upper limits from LoTSS-DR2 \label{tab:upperlimit_sources}}
  \centering
  \tablehead{
    \colhead{ID} &\colhead{SDSS Name} &\colhead{P} &\colhead{$z$} &\colhead{$F_{\rm p}$} &\colhead{$\log P_{\rm 20cm}$} &\colhead{$\log R$} &\colhead{$\log M_{\rm BH}$} &\colhead{$\log L/L_{\rm Edd}$} &\colhead{$\log L_{\rm H\alpha^B}$} &\colhead{$\log L_{5100}$}\\
    \colhead{} &\colhead{} &\colhead{} &\colhead{} &\colhead{mJy/beam} &\colhead{log(W/Hz)} &\colhead{} &\colhead{log(M$_\sun$)} &\colhead{} &\colhead{log(erg/s)} &\colhead{log(erg/s)} \\
    \colhead{(1)} &\colhead{(2)} &\colhead{(3)} &\colhead{(4)} &\colhead{(5)} &\colhead{(6)} &\colhead{(7)} &\colhead{(8)} &\colhead{(9)} &\colhead{(10)} &\colhead{(11)}
        }
  \startdata
  U1 & J074424.48+332034.7 & 1 & 0.0633 & <0.573 & <21.28 & <0.67 & 6.30 & -1.20 & 40.59 & 42.16\\
  U2 & J075038.12+315737.1 & 1 & 0.0548 & <0.417 & <21.02 & <0.77 & 6.00 & -1.30 & 40.16 & 41.79\\
  U3 & J075709.33+423616.4 & 1 & 0.0740 & <0.462 & <21.33 & <0.78 & 6.20 & -1.20 & 40.51 & 42.09\\
  \enddata
  \tablecomments{Column (1), identification number assigned in this paper for the upper-limit sources
  	(Prefix `U' means `Upper-limit').
  	Column (2) to (4), the same as in the above Table~\ref{tab:snr_gt3}.
  	Column (5), upper limit of the flux density at 0.15 GHz,
  	 set to be five times the rms value at the respective location of every source.
  	Columns (6) and (7), rest-frame 20 cm power and radio loudness derived with the values in Column (5).
  	Columns (8) to (10), virial mass of black hole, Eddington ratio and luminosity of broad H$\alpha$ emission line,
  	retrieved from the parent catalogs.
  	Column (11), continuum luminosity at rest-frame 5100 \AA\;
  	($L_{5100}\equiv \lambda L_\lambda \;{\rm at}\; \lambda=5100\;$\AA), derived from $L_{\rm H\alpha^B}$. \\
  (This table is available in its entirety in a machine-readable form in the online journal. A portion is shown here for guidance regarding its form and content.)}
\end{deluxetable}

\section{$\log N - \log S$ relation for the expanding, spatially flat Universe \label{appendix:logNlogS} }

Although the real Universe is likely flat for its 3-dimensional space with measured $k = 0$,
usually called \textit{spatially flat} or even \textit{spatially Euclidean} in the literature,
it is yet expanding or called time evolving (i.e., its 4-dimensional spacetime is curved).
Thus we must assess the impact of the cosmological expansion (namely redshift effects)
onto the \logNlogS/ test;
it concerns not only the measured frequency and fluxes
related to the so-called $K$-correction,
but also the geometrical relation between comoving volume (and thus density) and luminosity distance
(see \S11.1.1 of \citealt{1997iagn.book.....P} for the derivations).
\footnote{During our derivation we sense that
Equation (11.12) of \citet{1997iagn.book.....P} might be confusing and deserves a caution.
It is for a normal Euclidean universe,
	not for a spatially flat yet expanding universe,
	because that equation starts from
	the $r_\mathrm{max} - S$ relation of the non-evolving Euclidean case (their Equation 10.5).} 
Here we present the necessary formulation for the analyses in \S\ref{section:logNS}.

Generally for an expanding universe, the flux measured at some observer-frame frequency $\nu_0$
has the following relation with luminosity distance ($d_L$) or redshift ($z$):
\begin{equation}\label{eq:Fnu-dL}
F_\nu(\nu_0) = \frac{L_\nu(\nu_0)\, (1+z)^{1-\alpha}}{4\pi d_L^2} ~,
\end{equation}
where we assume the radio continuum is modeled as a power law, $L_\nu \propto \nu^{-\alpha}$.
On the other hand, for an isotropic universe,
the total number of sources per unit solid angle ($\mathrm{d}\Omega$)
detected above a specific limiting flux $S$ (at the observed frequency $\nu_0$)
can be expressed as the following volume integral:
\begin{equation}\label{N-S_definition}
N(S) = \int_V \frac{\mathrm{d}N}{\mathrm{d}\Omega} (F_\nu(\nu_0)\geqslant S)
  =\int_V \frac{\mathrm{d}N}{\mathrm{d}\Omega} (z\leqslant z_\mathrm{max}) ~,
\end{equation}
where $z_\mathrm{max}$ is the maximum redshift,
at which the sources with luminosity $L_\nu(\nu_0)$
can be detected at the limit $S$.
That is, $z_\mathrm{max}(L_\nu, S)$ satisfying Equation~\ref{eq:Fnu-dL}, which reads
\begin{equation}\label{eq:S-zmax}
S = \frac{L_\nu(\nu_0)(1+z_\mathrm{max})^{1-\alpha}}{4\pi d_L^2(z_\mathrm{max}) } ~,
\end{equation}
where $d_L(z_\mathrm{max})$ is the luminosity distance corresponding to $z_\mathrm{max}$,
the two being definitely linked as prescribed by the cosmological parameters.
Because of the complex formula of $d_L(z_\mathrm{max})$,
the relation between $S$ and $z_\mathrm{max}$ (Equation \ref{eq:S-zmax})
can be solved numerically only, and generally
the dependence of $S$ on distance deviates from
the simple inverse-square relation
of the non-expanding Euclidean universe.

For an spatially flat universe ($k=0$),
Equation~\ref{N-S_definition} can be expressed easily using comoving density and comoving volume.
Particularly, when a uniform comoving source density $n(r) = n_0$ is further assumed,
it becomes in a simple form
\begin{eqnarray}
N(S) & = & \int_0^{d_\mathrm{co}(z_\mathrm{max})} n(r) r^2 \mathrm{d}r \\
   & = & \frac{n_0}{3} d^3_\mathrm{co}(z_\mathrm{max}) ~. \label{N-Dco(max)}
\end{eqnarray}
Here $r$, $n_0$ and $n(r)$ are quantities with respect to the comoving frame
(see Page~157 of \citealt{1997iagn.book.....P} for the explanation),
and $d_\mathrm{co}(z_\mathrm{max})$ is the comoving distance
(namely proper distance at $z=0$) corresponding to $z_\mathrm{max}$,
and $d_\mathrm{co}(z_\mathrm{max}) = d_L(z_\mathrm{max}) / (1+z_\mathrm{max})$.

In order to demonstrate the complicated $N-S$ relation,
by combining Equations (\ref{eq:S-zmax}) and (\ref{N-Dco(max)}) we obtain
\begin{equation} \label{our_N-S}
N(S)  =  \frac{n_0}{3} (\frac{L_\nu(\nu_0)}{4\pi})^{1.5}\,
    (1+ z_\mathrm{max}(S, L_\nu)\/)^{-1.5\,(1+\alpha)}\, S^{-1.5} ~.
\end{equation}
Again, as discussed above, because $z_\mathrm{max}(S, L_\nu)$ is
a complicated function of $S$ (Equation \ref{eq:S-zmax}),
generally Equation (\ref{our_N-S}) deviates from $N \propto S^{-1.5}$, more or less.

\renewcommand{\thefigure}{D\arabic{figure}}
\setcounter{figure}{0}
\begin{figure}[htb]
  \centering
  \includegraphics[width=0.47\textwidth]{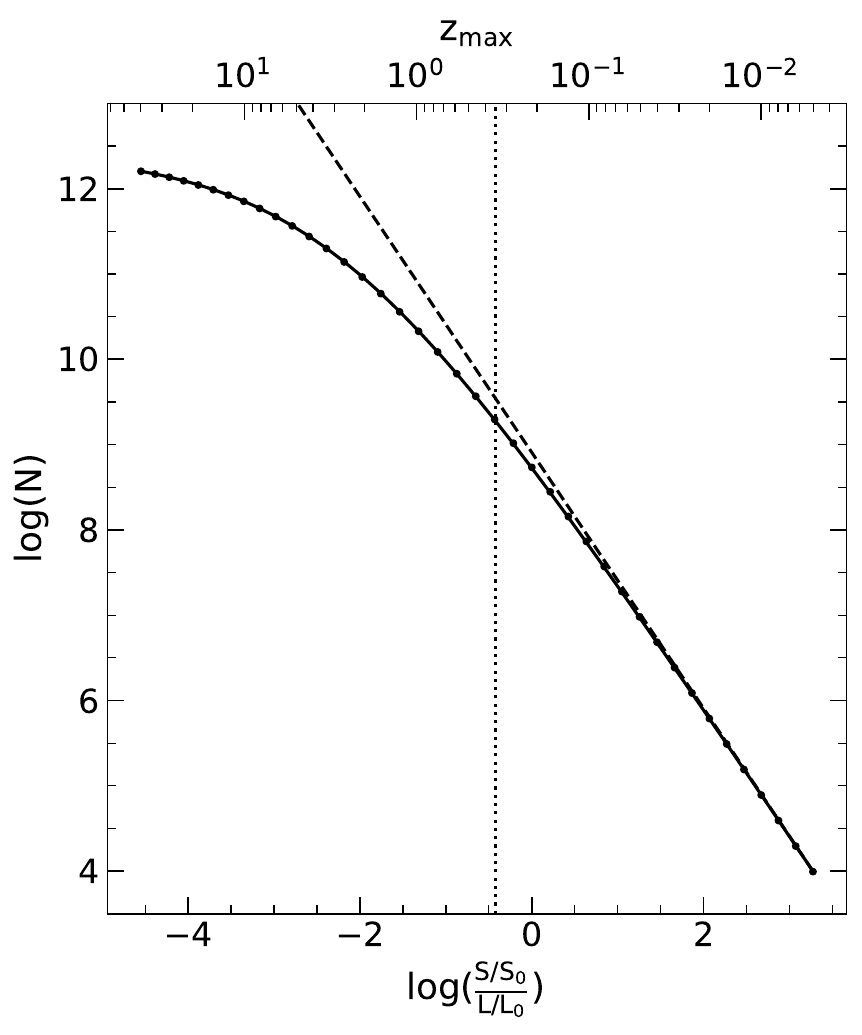}
  \caption{The numerically computed $\log N - \log \frac{S/S_0}{L/L_0}$ relation for the expanding spatially flat Universe.
  This relation (black dots) is in fact the $\log d_\mathrm{co}^3 - \log \frac{S/S_0}{L/L_0}$ relation
   with an arbitrary space density $n_0$.
  The canonical $\log N - \log \frac{S/S_0}{L/L_0}$ relation for the Euclidean universe (dashed line)
  is plotted just for comparison.
  The modified double power-law model, which is used to approximate the numerically computed relation,
  is also plotted (solid curve).
  The vertical dotted line indicates where $z_\mathrm{max} = 0.35$\,.}
  \label{fig:Append-logNS-expanding}
\end{figure}

We numerically solve the equation system (\ref{eq:S-zmax} and \ref{N-Dco(max)})
for the $\Lambda$CDM cosmology adopted in this work with the AGN continuum slope we used ($\alpha = 0.46$),
and obtain the relation between $\log N$ (or $\log d_\mathrm{co}^3$ equivalently) and $\log S$;
see the data points (solid circles) plotted in Figure~\ref{fig:Append-logNS-expanding}.

In fact, we use $\log N - \log \frac{S/S_0}{L/L_0}$ relation as displayed in the figure, after careful considerations.
Our primary consideration, as discussed in \S\ref{section:logNS}, is to keep the overall shape of the $N(S)$ curve
independent of the concrete values of $S$ and $L$ as well as $n_0$.
In other words, we do our best to achieve the following goal:
the effects of $S$, $L$ and $n_0$ to the curve shape
is just to shift the $N(S)$ curve vertically or horizontally in log--log scale.
Thus we make the following two manipulations.
(1) To use $S/L$ instead of $S$, which is in fact implied by Equation~\ref{eq:S-zmax}.
By doing so, in log--log scale the $N(S)$ curve only shifts horizontally
with changing $S$ or $L$, keeping its overall shape unchanged.
(2) To normalize $S$ and $L$.
The concrete values of the normalization factors $S_0$ and $L_0$, just like the above point (1),
do not impact the curve shape, and are somehow arbitrary in principle.
In practice, a good choice is to fix them to be the mean values of the data set under consideration,
because this choice is a natural rule of thumb of numerical calculation with computers,
and makes easy for us to incarnate a simple (and elegant) analytical form to approximate $N(S)$ (see below).
To make Figure~\ref{fig:Append-logNS-expanding},
we set $S_0$ and $L_0$ to be 1 mJy and $10^{22.57}$ W$\cdot {\rm Hz}^{-1}$, respectively,
adopt a single luminosity $L$ to be $L_0$ simply (see \S\ref{section:logNS}),
and set $n_0$ arbitrarily to be $3\;\mathrm{Mpc}^{-3}$;
the data points are calculated by scanning $z_\mathrm{max}$ (i.e., $\log \frac{S/S_0}{L/L_0}$ equivalently)
from $0.005$ to $50$.

Now comes the crux of this Appendix~\ref{appendix:logNlogS}:
construct an analytical function to approximate $N(S)$
in the way that, as stated in the above,
the curve shape can be used in any cases of arbitrary $S$, $L$ and $n_0$ values.
After Equation~\ref{our_N-S}, a modified double power-law model comes to our mind: 
\begin{equation}
  N = C\left(\frac{S/S_0}{L/L_0}\right)^{-1.5}\,
  \left[d_\mathrm{t} + \left(\frac{S/S_0}{L/L_0}\right)^{-\gamma}\right]^{\frac{1.5+\beta}{-\gamma}} ~.
  \label{eq:NS_doublePL}
\end{equation}
The first power law with the slope fixed to $-1.5$
is the only dominant component in the low-$z$ end (i.e., high-$S$),
which asymptotes to the Euclidean-universe case.
The second, modified power law component of this model
gets more and more significant as $S$ approaches toward the low end,
and works together with the first component to make the \logNlogS/ slope
get flatter and flatter and
asymptotic to 0 as $S$ tends to 0.
The characteristic parameter $d_\mathrm{t}$ accounts for the turnover of the curve,
and the exponent $\gamma$ controls the abruptness degree of the turnover.

We fit the modified double power-law model (\ref{eq:NS_doublePL}) to the numerically calculated data points
of the expansion version of the \logNlogS/ relation, and the fit is excellent
indicating the aptness of this carefully devised model
(see the solid line overplotted in Figure~\ref{fig:Append-logNS-expanding}).
The two normalization factors, $S_0$ and $L_0$,
are fixed to the mean values of the data set we used (see \S\ref{section:logNS}), as stated in the above.
In principle, $S_0$ and $L_0$ can be arbitrary;
but if so, the parameter $d_\mathrm{t}$ would not be regarded as invariant,
yet would have to be shifted by a quantity like $(S_0/L_0)^\gamma$
when applying the above model to any data set.
The parameters $C$ is a free parameter to be fitted to match $N$.
The best-fitted values of the three shape parameters, $\beta$, $\gamma$ and $d_\mathrm{t}$,
are listed in \S\ref{section:logNS} already.
Here we explain once again that we can directly apply the best-fitted $d_\mathrm{t}$ value to
\S\ref{section:logNS}, because we use the same $S_0$ and $L_0$ here and there.

We also plot the canonical $\log N - \log \frac{S/S_0}{L/L_0}$ relation
(namely $N \propto S^{-1.5}$, the static version)
in Figure~\ref{fig:Append-logNS-expanding} as comparison,
with the same $S_0$, $L_0$, $L$ and $n_0$ values.
Compared with the canonical relation, the expansion-modified version
differs little in the bright $S$ end
(i.e., for the cases with very small $z_\mathrm{max}$; see Equation~\ref{eq:S-zmax}) as expected,
yet gets flatter (namely with less and less sources) toward the faint $S$ end (i.e., for higher $z_\mathrm{max}$).
In the case of $z_\mathrm{max} = 0.35$ that is
the maximum redshift of our parent sample (the optical low-mass AGNs),
for the same flux limit ($S$) the expansion-modified version
predicts 0.52 times the number ($N$) of the canonical relation.
This deviation is too significant to be neglected.


\end{document}